\documentclass{cernyrep} 
\usepackage{texnames}
\usepackage[T1]{fontenc}
\usepackage[colorlinks=true, pdftex]{hyperref}

\begin{document}
  \title{Beyond the Standard Model}
\author{A. Pomarol}
\institute{Universitat Aut\`onoma de Barcelona, Bellaterra, Spain}
\maketitle 

\begin{abstract}
In these lectures  we briefly cover   some of the main lines of research in particle physics beyond the Standard Model
\end{abstract}
 
\section{Introduction}

The Standard Model (SM) of elementary particles
together with   Einstein's theory of General  Relativity for gravity  
provide us  with  a  remarkable, simple  framework  to  explain, at present, almost all
physical processes observed in Nature.
Only physics at shorter distances than the Planck length, where gravity must be 
fully quantized, and
 some experimental evidence  (neutrino masses, dark matter)
seem to escape from this general framework and require us to go beyond it.
This new physics, however, could  appear  only  at around Planckian energies, thus
not providing strong motivations  for new feasible experiments  testing
physics at smaller energies such as the LHC.

A different  stimulus for   physics beyond the SM,
that has inspired a wide range of experiments,
has originated from  trying to  improve  our  theoretical   understanding  of  the SM. 
Indeed, certain  couplings and masses in the SM,  determined by experiments,
seem to  demand  a better explanation, and this has  required us to  postulate new physics beyond the SM.
 The purpose of these lectures is  to give  a brief description of these theories,
giving  their motivation and predictions at present and future experiments.
We have divided  the lectures  into the following topics:
\begin{itemize}
\item The SM of particle physics: symmetries, consistency, and
reasons for improvement.
\item Grand Unified Theories.
\item The strong CP-problem and axions.
\item The hierarchy problem.
\item Supersymmetry.
\item Higgsless models and composite Higgs.
\item Extra dimensions.
  \end{itemize}
 
\section{The  SM of particle physics: symmetries, consistency, and
reasons for improvement}

The SM is  a  quantum field theory whose Lagrangian, that  gives us the particle spectrum and interactions, is   fixed by  local symmetries  and 
the matter content. 
The local symmetries of the SM  are those associated to
\begin{enumerate}
\item The Poincar\'{e} group.
\item The gauge group $SU(3)_c\times SU(2)_L\times U(1)_Y$.
\end{enumerate}
The matter content, defined by  the quantum numbers under the  above groups,
consists of a  fermionic and a scalar sector.
The fermionic sector  is  composed of three copies of   fields charged
under the gauge group\footnote{We are using  the convention of charges such that the EM charge is defined as $Q=Y/2+T_3$
where $T_3$ is the 3rd component of the $SU(2)_L$ generator and $Y$ is the charge under the $U(1)_Y$ group, the so-called hypercharge.}:
\begin{equation}
\begin{array}{c|c|c|c} & SU(3)_c & SU(2)_L & U(1)_Y \\\hline Q_L & 3 & 2 & 1/3 \\\hline u_R & 3 & 1& 4/3 \\\hline d_R & 3 & 1 & -2/3 \\\hline l_L & 1 & 2 & -1 \\\hline e_R & 1 & 1 & -2\end{array}
\end{equation}
Each copy corresponds to, what  is usually called, a family of particles.
They are fields of spin $1/2$.
The scalar sector contains  the Higgs. This is a scalar with  charges:
\begin{equation}
\begin{array}{c|c|c|c} & SU(3)_c & SU(2)_L & U(1)_Y \\\hline H & 1& 2 & 1 
\end{array}
\end{equation}
Invariance under the gauge symmetries  requires  the presence 
of additional gauge boson fields, the gluons $G_\mu$, the $W_\mu$ and $B_\mu$.
These are fields of spin 1.
 Finally, invariance under the Poincar\'{e} group requires the presence 
 of the gravitational field $g_{\mu\nu}$ of spin 2.
 All these fields mediate interactions between matter fields.
 
Once the local symmetries and the matter content of the SM are fixed, the Lagrangian
is fully determined.  Neglecting for the moment gravity whose strength is very small compared to 
other  interactions in particle physics experiments,  we have that
the renormalizable SM Lagrangian is given by
\begin{eqnarray}
{\cal L}_{\rm SM}&=&-\frac{1}{4g'^2}B^{\mu\nu}
B_{\mu\nu}
-\frac{1}{4g^2}W^{\mu\nu}_a
W_{\mu\nu}^a
-\frac{1}{4g_s^2}G^{\mu\nu}_a
G^a_{\mu\nu }
\nonumber\\
&+&i\bar Q_L^i\,\slash \hskip -.25cm DQ_L^i
+i\bar u_R^i\, \slash \hskip -.25cm Du_R^i
+i\bar d_R^i\,\slash \hskip -.25cm Dd_R^i
+i\bar e_R^i\,\slash \hskip -.25cm De_R^i 
+i\bar l_L^i\,\slash \hskip -.25cm Dl_L^i\nonumber\\
&+&Y_u^{ij}\bar Q_L^i  \tilde H u^j_R+
Y_d^{ij}\bar Q_L^i  Hd_R^j+
Y_e^{il}\bar l_L^i  He_R^j+h.c.\nonumber\\
&+&|D_\mu H|^2
+\mu^2 |H|^2-\lambda  |H|^4\, ,
\label{sml}
\end{eqnarray}
where   $i,j=1,2,3$ label the families.
The explanation for each term of this Lagrangian is given in the lectures of A. Pich
in this report.
Apart from  kinetic terms and mass terms, this Lagrangian gives us the interactions
between the SM particles. We  have gauge interactions whose strengths are measured by $g_s$, $g$ and $g'$, Yukawa interactions $Y_{u,d,e}$, and Higgs self-interations measured by $\lambda$.
 The important feature of the SM is that all these couplings are dimensionless (in natural units $\hbar=c=1$) and therefore  the theory can be extrapolated into a wide range of energies.
We will discuss this in more detail later on.

By rotating  the  three families of fermion fields, 
we can  go to the basis in which the Yukawa couplings can be written as 
\begin{equation}
Y_d=V_{\rm CKM}^\dagger M_d^{\rm diag}/\langle H\rangle\ ,\ \ \ 
Y_u=M_u^{\rm diag}/\langle H\rangle\ ,\ \ \
Y_e= M_e^{\rm diag}/\langle H\rangle\,,
\end{equation}
 where  $M_{d,u,e}^{\rm diag}$ are the diagonal fermion mass matrices, $V_{\rm CKM}$ is the Cabibbo-Kobayashi-Maskawa matrix,
and $\langle H\rangle$ is the Higgs vacuum expectation value (VEV).

We have measured all parameters of the SM \cite{Nakamura:2010zzi}, except the Higgs self-coupling $\lambda$
that determines the Higgs mass $M_h=\sqrt{2\lambda}v$ where $v=\sqrt{2}\, \langle H\rangle\simeq 246$ GeV.
Direct Higgs searches at LEP and the Tevatron, and 
electroweak  precision tests (EWPT) give important constraints  on the Higgs mass.
One gets $114.4$ GeV$<M_h<155$ GeV  at 95$\%$ CL \cite{gfitter}.

\subsection{Accidental symmetries}

The Lagrangian Eq.~(\ref{sml}) has extra symmetries that appear  'accidentally'
since we did not impose them. These symmetries allow one to explain
certain properties of the SM.
The most important one is  baryon number. This is  a global $U(1)_B$  symmetry under which the SM fermions $\psi$ transform as
\begin{equation}
\psi\rightarrow e^{iB\theta}\psi\, ,
 \label{bs}
\end{equation}
where $\theta$ is the parameter of the transformation and $B$ is the baryon number.
We have $B=1/3$ for the quark fields $Q_L$, $u_R$ and $d_R$,  while $B=0$ for leptons.
It is not hard  to see that, indeed, this is a global symmetry of  Eq.~(\ref{sml}). 
This accidental symmetry   has an important implication.
At the perturbative level\footnote{This symmetry is, however, `anomalous', {i.e.}  broken at the quantum level. This implies that the SM predicts that the proton should decay but at an extremely
small rate.},  it guarantees that the proton, that is made of three quarks and then has baryon number $B=1$, cannot decay into lighter leptonic states. 
This prediction of the SM  is supported by the experimental data that, up to now,  has not shown any evidence for proton decay \cite{Nakamura:2010zzi}.

Other accidental symmetries of the  SM  Lagrangian  are  the three leptonic global symmetries. 
These are  (1) the electronic lepton symmetry, whose charges are  $L_e=1$ for electrons and the neutrino $\nu_e$ and zero
for the rest;
(2) the muonic lepton symmetry, whose charges are  $L_\mu=1$ for muons and  the neutrino $\nu_\mu$
and zero
for the rest;  and 
(3) tauonic lepton symmetry, whose charges are  $L_\tau=1$ for taus and the neutrino $\nu_\tau$ and zero for the rest.
These symmetries forbid  leptons to  decay into a lighter  one plus a photon, e.g.,
 $\mu\rightarrow e\gamma$.
Again, this prediction is, at present, supported by experimental data that has  shown no   evidence for these decays \cite{Nakamura:2010zzi}.
It also predicts that neutrinos cannot have mass, in clear  contradiction with experimental
evidences of neutrino oscillations. We will comment on this point later.

The SM Lagrangian also has {\it approximate} accidental symmetries
that play an important role in the  understanding of  some physical properties of the model.
These are global symmetries that are only broken by small couplings in the SM.
An important one is the so-called `custodial' $SU(2)_c$ symmetry \cite{Sikivie:1980hm}:
in the limit $Y_{u,d,e}=0$ and $g'=0$, the SM Lagrangian has an extra   global  
$SU(2)$ symmetry
under which the Higgs field $H$  transforms\footnote{This is a $SU(2)$ rotation between    $H^r$ and $\epsilon^{rs} H_s$ where $r,s=1,2$ label the components of a   $SU(2)_L$ doublet.} as a $\bf 2$.
The Higgs VEV breaks this symmetry down to the custodial symmetry, $SU(2)\times SU(2)_L\rightarrow SU(2)_c$, under which
the physical Higgs  $h$ transform as a  singlet and the massive $W^\pm_\mu$ and $Z_\mu$  form a triplet. This symmetry  predicts $M_W=M_Z$.
When $g'$ is turned on, this prediction is altered  to be 
\begin{equation}
\frac{M^2_W}{ M^2_Z\cos^2\theta_W}\equiv \rho =1\, ,
\label{rhop}
\end{equation}
where  $\theta_W$ is the weak angle.
Yukawa couplings also  break the custodial  symmetry and  therefore  modify  the above prediction. Nevertheless, these effects arise at the one-loop   order  and give modifications of order $Y_t^2/(16\pi^2)\sim 0.01$  at most ($Y_t$ is the top Yukawa coupling). 
Once more, an accidental symmetry, in this case not exact but approximate, gives a nontrivial
prediction, $\rho\simeq 1$, that has been very well tested by experiments. 

Other important  approximate  accidental symmetries of the SM are flavour symmetries.
In the limit of $Y_{d,u,e}=0$, the SM has five extra global $SU(3)$ symmetries associated to unitary transformations
of the three families of $Q_L^i$, $u_R^i$, $d_R^i$, $l_L^i$ and $e_R^i$ respectively. 
These symmetries, or subgroups of them, allow us to explain a plethora  of 
  flavour physics  properties,  as you can find  in the lectures of  G.~Hiller in this report.

All the accidental symmetries  described above 
are symmetries of the SM due to the fact that 
  only  operators of  dimension $\leq 4$  are included in Eq.~(\ref{sml}).
As soon as we consider terms involving  higher dimensional operators  (terms having more fields or derivatives than those in Eq.~(\ref{sml})),  these accidental symmetries  are no longer preserved.
For example, we can write the dimension-five operator
\begin{equation}
{\cal L}_{\rm SM}^{(5)}=\frac{c_5}{\Lambda}\bar l^{c}_L \cdot H\,  H\cdot l_L+h.c.\, , 
\label{dim5}
\end{equation}
where $A\cdot B=\epsilon_{rs}A^rB^s$ with $r,s=1,2$ being $SU(2)_L$ indices, and
 $l^c_L$ is the conjugated of $l_L$.
This term is suppressed by a mass scale $\Lambda$ since the Lagrangian has to have dimension $4$ (the  coefficient $c_5$ is defined to be dimensionless).
One can easily realize that the term in Eq.~(\ref{dim5}), present in principle for each SM lepton,
violates  the lepton symmetries.
There are also operators of dimension 6 that violate  $L$ and $B$.
In particular
\begin{equation}
{\cal L}_{\rm SM}^{(6)}=\frac{c_6}{\Lambda^2}\epsilon_{rs}\epsilon_{\alpha\beta\gamma}[
\bar Q^{c\, r\, \alpha}_{L}\gamma^\mu u_{R}^{\beta}]
[\bar d^{c\, \gamma}_{R}\gamma_\mu l_{L}^s]+h.c.+\cdots\, 
\label{dim6}
\end{equation}
where $\alpha,\beta,\gamma=1,2,3$ run over the colour indices of $SU(3)_c$,
$r,s=1,2$ are $SU(2)_L$ indices,
and $c_6$ is a dimensionless coefficient that
violates not only the lepton symmetries but also  baryon number.
If present in the SM, it will lead to proton decay.
The experimental evidence of $B$ and $L$ conservation 
tells us that these terms must be either absent or be highly suppressed, {i.e.}, $\Lambda\gg M_W$.

\subsection{Consistency  of the SM}

There are other reasons for avoiding    terms such as those of Eqs.~(\ref{dim5}) and (\ref{dim6})
in the SM Lagrangian. We do not know how to quantize 
a field theory with operators of dimensions larger than 4.
The best we can do in the presence of these higher dimensional operators
 is to assume that
 the SM is an effective field theory valid only up to energies  $E\lesssim\Lambda$.
Theorists call this scale $\Lambda$, that determines  the energy  below  which a theory is valid, the `cutoff' scale.

A  question that immediately  comes  to mind is the following. 
Can  we,  in the   SM,   take  $\Lambda\rightarrow\infty$?
Or, in other words, is the SM  a theory valid for all energies?
We can  try to answer this question  either theoretically or experimentally.
In the first case,  we  must  check whether   the SM predicts always consistent results. 
To do so, we perform   Einstein's  Gedankenexperiment (thought experiment)
looking for inconsistencies of the SM, similar to those that led Einstein  to
postulate the theory of Special Relativity in order  to  reconcile   the laws of Newtonian classical mechanics with the laws  of electromagnetism.
Experimentally, we can also address the above question
by asking   whether  the SM  Lagrangian of Eq.~(\ref{sml})
explains all present experimental data. If not,   an extension of it will be needed
implying that the SM has a finite cutoff scale $\Lambda$ above which new physics shows up.

\begin{figure}
\centering\includegraphics[width=.75\linewidth]{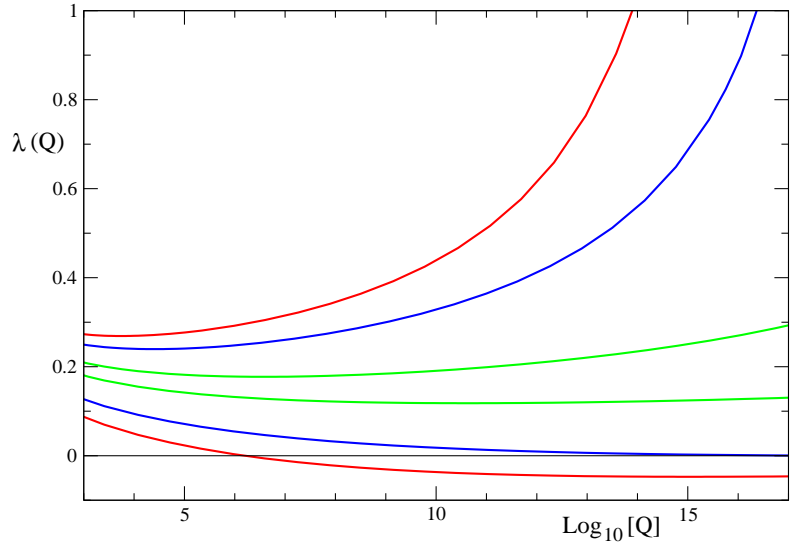}
\caption{Value of $\lambda(Q)$ as a function of $Q$  for different initial values of $\lambda(Q=10^3$ GeV)}
\label{plotlambda}
\end{figure}

\begin{figure}
\centering\includegraphics[width=.75\linewidth]{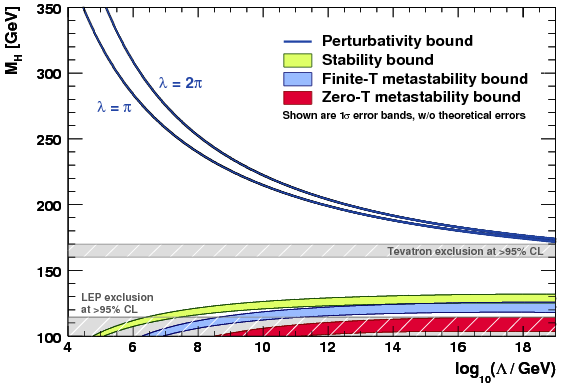}
\caption{Values of $M_h$ as a function of $\Lambda$. See details in Ref.~\cite{Ellis:2009tp}
.}
\label{higgsmrange}
\end{figure}

\begin{figure}
\centering\includegraphics[width=.5\linewidth]{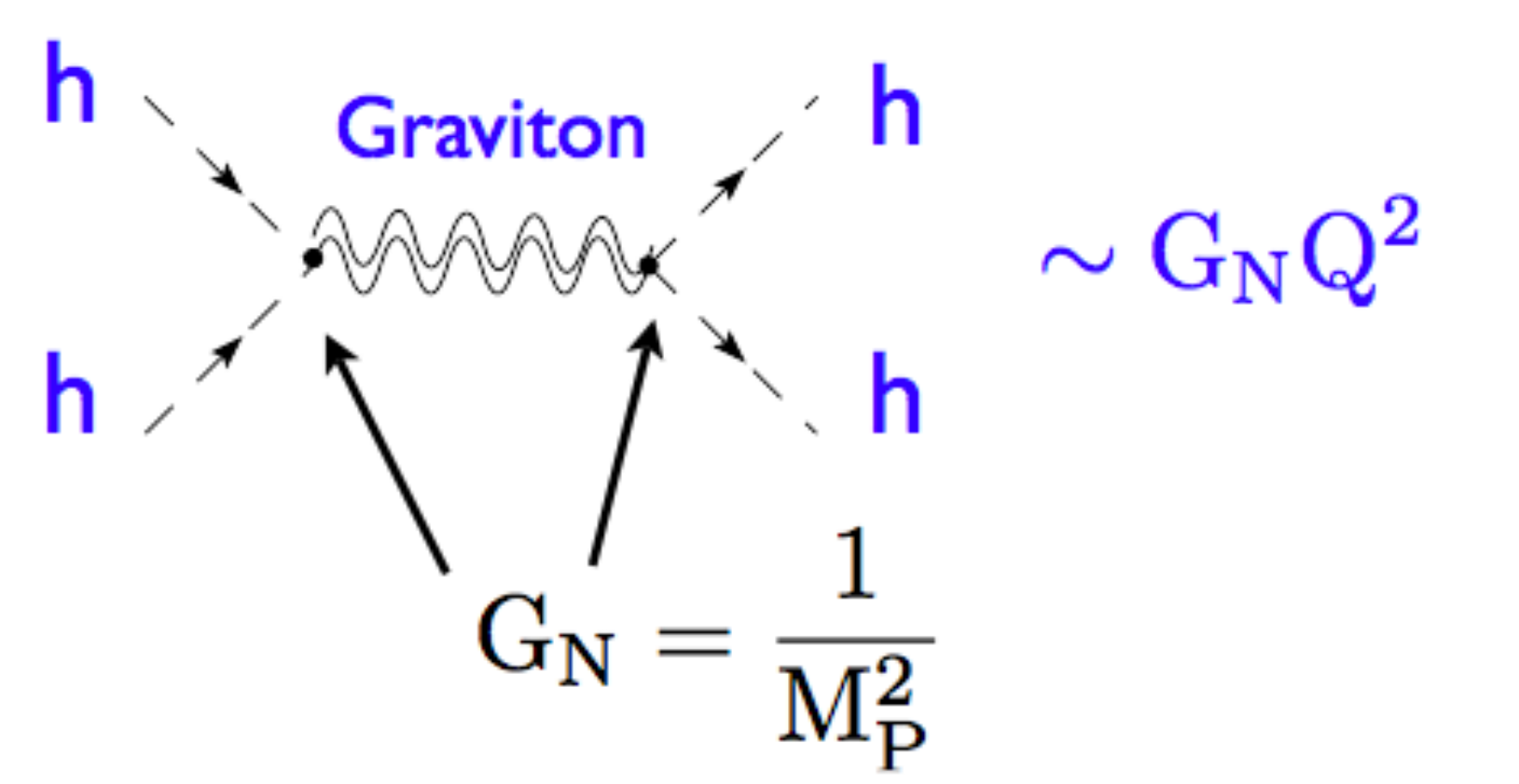}
\caption{Feynmann diagram for the contribution of a graviton to the $hh\rightarrow hh$ process}
\label{gravity}
\end{figure}

Let us first  try  to answer the above question from the theoretical perspective.
The   Gedankenexperiment that we  proposed here, in order to check the validity of the SM,
 is to calculate the   amplitude of Higgs scattering, 
 $hh\rightarrow hh$, at very high energies.
 At tree level this amplitude is proportional to $\lambda$.
 At the quantum level this amplitude is well approximated by using  the
  `running coupling' $\lambda(Q)$,
where   the scale $Q$ can be approximately  associated with the energy at the centre of mass of the process.   The running coupling $\lambda(Q)$ can be easily obtained 
 by solving the renormalization group equation (RGE):
 \begin{equation}
\frac{d\lambda (Q)}{d\ln Q}=\frac{1}{16\pi^2} (24\lambda^2 +12\lambda Y^2_t-6Y^4_t)+\cdots\, ,
  \end{equation}
  where we only show the dominant one-loop result arising from the top and Higgs. 
 As we increase $Q$, $\lambda(Q)$ can increase or decrease, depending on its initial value, taken, for example, at $Q=10^3$ GeV. 
This is shown in   Fig.~\ref{plotlambda}. If $\lambda(Q)$ grows with $Q$, we can reach a point at which $\lambda(Q)$ is too large and we cannot calculate within perturbation theory.
Let us call the  scale at which this happens $\Lambda_+$.
At $Q=\Lambda_+$ the SM becomes intractable, and  even lattice studies (used to study  theories with large couplings   such as QCD) have shown  that the SM cannot take such initial values of $\lambda$ if at the same time we demand  $\Lambda_+\rightarrow \infty$.
In other words,   for those  initial  values of  $\lambda$  for which  
 $d\lambda(Q)/d\ln Q>0$, the SM 
is only valid  up  to  energies $\sim\Lambda_+$, {i.e.}, the SM has a nonzero cutoff scale $\Lambda\sim \Lambda_+$.
On the other hand, if $\lambda(Q)$ decreases with $Q$, it can become negative  
at some $Q=\Lambda_-$  in which  case the electroweak  vacuum
is only a local minimum  and  there is a new deeper and potentially dangerous minimum
at this scale. 
Up to some caveats discussed in Ref.~\cite{Ellis:2009tp}, the SM cannot be  valid at
energies at which $\lambda(Q)$ is negative, showing  again  a  limitation in  the SM, $\Lambda\sim \Lambda_-$. 
By relating $\lambda$ with the Higgs mass, $M_h=\sqrt{2\lambda(Q\sim M_h)}v$, we show in Fig.~\ref{higgsmrange} the range of validity of the SM  (the scale $\Lambda$)
versus  $M_h$.
We can see that there is only a small window  around  $M_h\sim 150$ GeV in which
the SM is  valid, at least,   up to $Q \sim 10^{19}$ GeV.  
Why do we stop at this scale?
Because at energies  above   $10^{19}$ GeV, gravity becomes important and must be considered
in the calculation. The new contribution,  given in Fig.~\ref{gravity},
grows with $Q$
as $G_N Q^2$, where $G_N$ is the Newton constant, and   becomes  relevant at energies $\sim \sqrt{1/G_N}\equiv M_P\simeq 10^{19}$ GeV. 
We do not know how to treat at the quantum level a theory of gravity
at energies above the Planck scale $M_P$.
For this reason the SM has  an ultimate cutoff\footnote{One can check that the other SM interactions are always well behaved at energies below $M_P$.}  scale $\Lambda\sim M_P$. 
Above this scale we do not know how to calculate physical quantities. 
It is interesting to compare this situation with Fermi's  theory.
When Fermi proposed his theory for the weak interactions it was clear for him that this was a theory valid up to energies  $\sim \sqrt{1/G_F}\simeq 300$ GeV.  He was right, and we know today that at energies  around  $\sqrt{1/G_F}$ there is `new physics'
beyond Fermi theory, the $W$ and the $Z$ gauge bosons.
Similarly,  we expect that  at energies around $\Lambda\sim M_P$ 
new physics beyond the SM will show up. 
 One possibility for  this new physics is string theory that consists in 
 replacing fields and particles by strings. The SM particles would correspond to massless
 string excitations, while massive excitations will have masses of order $M_P$.
 
Let us now move to the experimental consistency of the SM.
Is there any experiment that cannot be explained by the SM?
We find four pieces of experimental evidence beyond the SM:
\begin{enumerate}
\item Neutrino oscillations that require that neutrinos be massive.
\item The need for dark matter  in the Universe.
\item  The presence of a cosmological inflationary epoch.
\item The matter-antimatter asymmetry in the Universe.
\end{enumerate}
Indeed,  the SM predicts  that  
neutrinos are massless,  does  not have a  candidate for dark matter,
and  a  cosmological inflationary epoch  or 
 a matter--antimatter asymmetry cannot be produced  
    at the desirable rate.
Although physics beyond the SM is required, it is important to 
be aware  of the fact that  all    the above  experimental evidence beyond the SM 
does not really require new physics at  scales much below $M_P$.
Then,  since as explained above the presence of gravity tells us 
that the  SM is an effective  theory with a maximum cutoff scale around $M_P$,
we can expect Planckian physics  to be  responsible for  the above experimental disagreements. 
For example, if  the SM is not valid above  $\Lambda$, we can expect terms as those
of Eqs.~(\ref{dim5}) and (\ref{dim6}) to be present in the theory.
After electroweak symmetry breaking (EWSB),  Eq.~(\ref{dim5}) leads to neutrino masses  of order
\begin{equation}
m_\nu\sim\frac{v^2}{\Lambda}\sim  0.06\ {\rm eV}\ \left(\frac{10^{15}\ {\rm GeV}}{\Lambda}\right)\, .
\end{equation}
Therefore  neutrino masses can be induced with  the right magnitude for a  $\Lambda$  not far away from  $M_P$.  Also the other experimental conflicts can be blamed on physics
at around the Planck scale, so one can maintain that experiments do  not really provide any evidence for  a lower cutoff scale in the SM.

\subsection{Reasons for improvement in the SM}

So far we have seen that the SM can be a theory valid all the way to energy  scales
around the Planck scale.
Experimental or theoretical inconsistencies are not enough to 
put the SM in trouble at lower energies.
There are, however,  other theoretical reasons 
to go beyond the SM:
the  search for a `natural' explanation of the SM parameters.
Here we list some of them, from the most important one to the less important according to my taste:
\begin{enumerate}
\item {\it The cosmological constant:} \\
Experimentally we know that it takes the value  
$\Lambda_{cosmo}\sim  10^{-47}$ GeV$^4$.\\
 Theoretically we would have expected $\Lambda_{cosmo}\sim\Lambda^4\sim M_P^4\sim 10^{76}$ GeV$^4$.
\item {\it The Higgs mass term:}\\
  In order to give the right Higgs VEV, we must have $\mu^2\sim v^2\sim 10^{4}$ GeV$^2$.\\ Theoretically one would have expected 
$\mu^2\sim\Lambda^2\sim M_P^2\sim 10^{38}$ GeV$^2$.
\item {\it Charge quantization:}\\
We do not have any explanation  in the SM of why the electron charge is equal but opposite in sign to the proton charge, as experiments suggest to us: $Q_e+Q_p<10^{-21}$.
\item {\it The strong CP problem:}\\
We do not understand why in the SM the term   $\int d^4x\, \theta\, \epsilon^{\mu\nu\rho\sigma}G^a_{\mu\nu}G_{\rho\sigma}^a$, that would lead to CP violation in the strong sector, has a coefficient $\theta$ so small. Experimentally we know that $\theta\lesssim 10^{-13}$.
\item {\it Fermion masses and mixing angles:}\\
The fermion mass spectrum ranges from $\sim 170$ GeV, for the case of the top-quark, to $\sim 10^{-3}$ GeV, for the case of the electron. We do not know why there exists such  a large difference
in masses. We also  find   experimentally  that the Cabibbo--Kobayashi--Maskawa matrix is very close to a diagonal matrix. We do not know why.
\item {\it Gauge coupling unification:}\\
 We find experimentally that $g_s\sim 1.12$, $g\sim  0.65$, and 
$g^\prime\sim 0.35$ at $E\sim M_Z$. Is there any reason for such differences?
\item {\it Number of families:}\\  Matter is made  of  three families. Is there any reason for this?
\end{enumerate}
Let us briefly comment on them.
We do not have any natural explanation for  the cosmological constant value; it is still a true mystery  to us.
We do, however, have several possible   explanations  for  the smallness   of the Higgs mass as compared to   $M_P$  (usually referred as the hierarchy problem). Below
   we will discuss the most  interesting ones,
supersymmetry, Higgsless models, composite Higgs, and extra dimensions.
On charge quantization and gauge coupling, unification theorists have postulated the
existence   of   Grand Unified Theories   at high energies 
that could explain these relations.
We will discuss  them   below.
For the  strong CP problem, we have a nice explanation with a nice prediction, the axion state.
We will devote Section~\ref{stroncpproblem} to it.
Finally, several explanations for the observed fermion masses, mixings, and the number of families  have been proposed in the literature in the last years. I do not find any of them  compelling enough to single them out here. The usual problem with 
models of fermion masses  is that they do not lead to  sharp predictions.

\section{Grand Unified Theories (GUT)  \cite{Langacker:1980js}}

If we open the 2010 edition of the Particle Data Group \cite{Nakamura:2010zzi} we find
\begin{equation}
|Q_e+Q_p|/e<1.0\times 10^{-21}\, .
\end{equation}
This strong constraint suggests  that the electric and proton charge are quantized following the relation
\begin{equation}
Q_e=-Q_p\, .
\label{quant}
\end{equation}
Since in the SM we have $Q=Y/2+T_3$,  Eq.~(\ref{quant})  implies that the hypercharges are quantized following the relations
\begin{equation}
Y_{l_L}=2Y_{e_R}=-\frac{4}{3}Y_{u_R}=\frac{2}{3}Y_{d_R}=-\frac{1}{3}Y_{Q_L}
\, .
\label{quanty}
\end{equation}
In the SM, since the $U(1)_Y$ is an Abelian group,  the   hypercharges  could have been, in principle,  arbitrary real numbers. It is then surprising to find the relation Eq.~(\ref{quanty})\footnote{We must say, however, that the SM hypercharges are not really free parameters since  the absence of quantum anomalies in the SM forces them to fulfil a set of equations.
Equation~(\ref{quanty})  is a particular  case that leads to  an  anomaly-free theory.}.

A possible explanation for the SM hyperchage quantization comes from   assuming that the  SM group of symmetries is, at high energies, a much larger  group $G$.
If this group $G$ only contains non-Abelian groups all charges will be quantized.
The minimal group $G$ fulfilling this requirement is   the Pati--Salam group   \cite{Pati:1974yy}
$SU(4)\times SU(2)_L\times SU(2)_R$. 
Demanding $G$ to be a simple group, the minimal case is $G=SU(5)$, the model of Georgi and Glashow \cite{Georgi:1974sy}.
We discuss it next.

\subsection{$SU(5)$ GUT}

The $SU(5)$ group 
 is defined as the set of $5\times 5$ unitary matrices with determinant 1.
 This group
contains  $SU(3)\times SU(2)\times U(1)$ as a subgroup,  corresponding respectively to the $5\times 5$ matrices
\begin{equation}
\left(\begin{array}{c|c}  U_{3\times 3}   &  \\\hline & 0  \end{array}\right)\qquad ,\qquad
\left(\begin{array}{c|c}  0   &  \\\hline & U_{2\times 2}  \end{array}\right)\qquad ,\qquad
\left(\begin{array}{ccccc} e^{i\frac{2}{3}\theta} &  &  &  &  \\ & e^{i\frac{2}{3}\theta} &  &  &  \\ &  & e^{i\frac{2}{3}\theta} &  &  \\ &  &  & e^{-i\theta} &  \\ &  &  &  & e^{-i\theta}\end{array}\right)\, ,
\end{equation}
where $U_{3\times 3}$ ($U_{2\times 2}$) is a $3\times 3$ ($2\times 2$) matrix. 
The last matrix, to be associated with  a $U(1)_Y$ transformation,  shows
that, as expected,  the hypercharges are no longer free numbers  but they are quantized.
The $SU(5)$ group has  24 generators, each of them  has an associated  gauge boson.
Only 12 of then can be identified with the SM gauge bosons.
The other 12 are extra gauge bosons that must get masses $M_{\rm GUT}$ above the electroweak scale
where the $SU(5)$ must be broken.
These extra gauge bosons, referred to as $X$ and $Y$ bosons, have charges ${\bf (3,2)_{ 5/3}}$ and ${\bf (\bar 3,2)_{- 5/3}}$ under
the SM group.

\begin{figure}
\centering\includegraphics[width=.65\linewidth]{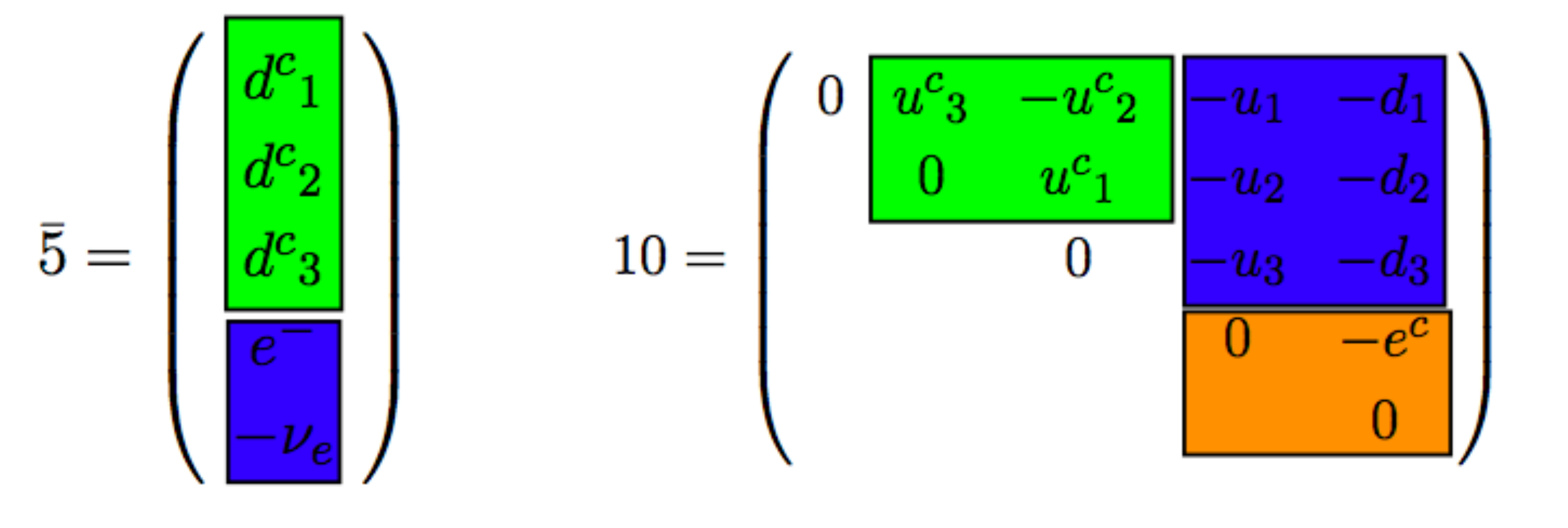}
\caption{The embedding of a family of  SM fermions. The indices $1,2,3$ refer to the three colours.
The fields $d^c$, $u^c$, and $e^c$  are the conjugated of $d_R$, $u_R$, and $e_R$ respectively.}
\label{su5}
\end{figure}

The SM fermions must be embedded in $SU(5)$ representations.
Amazingly, Georgi and Glashow realized that a full SM family of fermions could be neatly embedded into two $SU(5)$ representations, the ${\bf  \bar 5}$ and the ${\bf 10}$. 
The explicit  embeddings are given in Fig.~\ref{su5}.
These embeddings give the correct hypercharge assignments  Eq.~(\ref{quanty}).
Such simplicity, however, does not occur in the embedding of  the Higgs doublet into an $SU(5)$ representation. The minimal case is to  
embed  the Higgs into a ${\bf  5}$,  but this requires one to introduce a  colour triplet
 accompanying the Higgs.  Similarly to the  $X,Y$ bosons, this colour triplet must get a  mass when $SU(5)$ is broken.  This is known as the doublet-triplet splitting problem in GUT.

\begin{figure}
\centering\includegraphics[width=.7\linewidth]{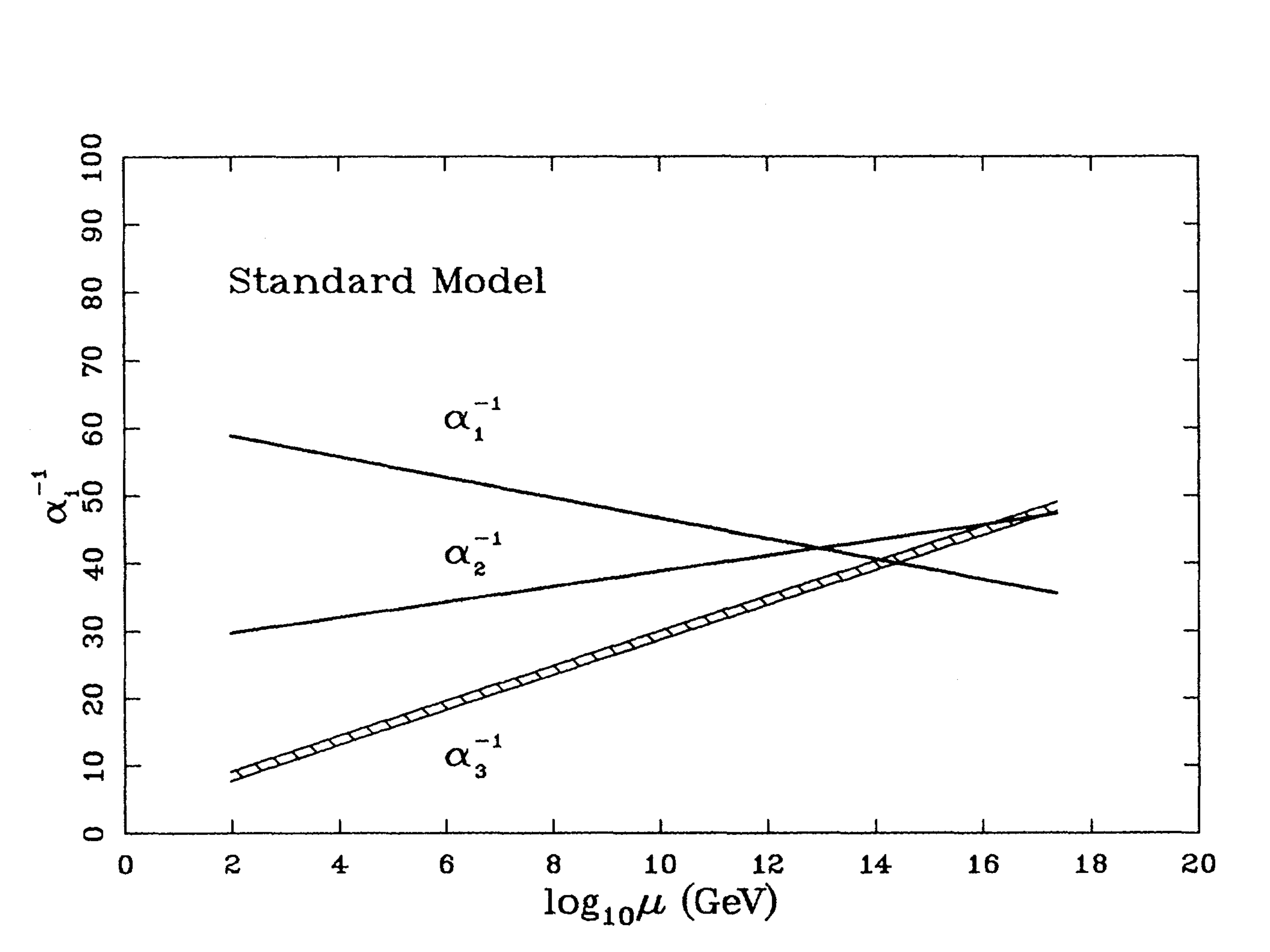}
\caption{Evolution of the three  SM gauge couplings $\alpha_i=g_i^2/(4\pi)$
as a function of $\mu=Q$ in the SM \cite{Langacker:1992rq}}
\label{gusm}
\end{figure}

\begin{figure}
\centering\includegraphics[width=.7\linewidth]{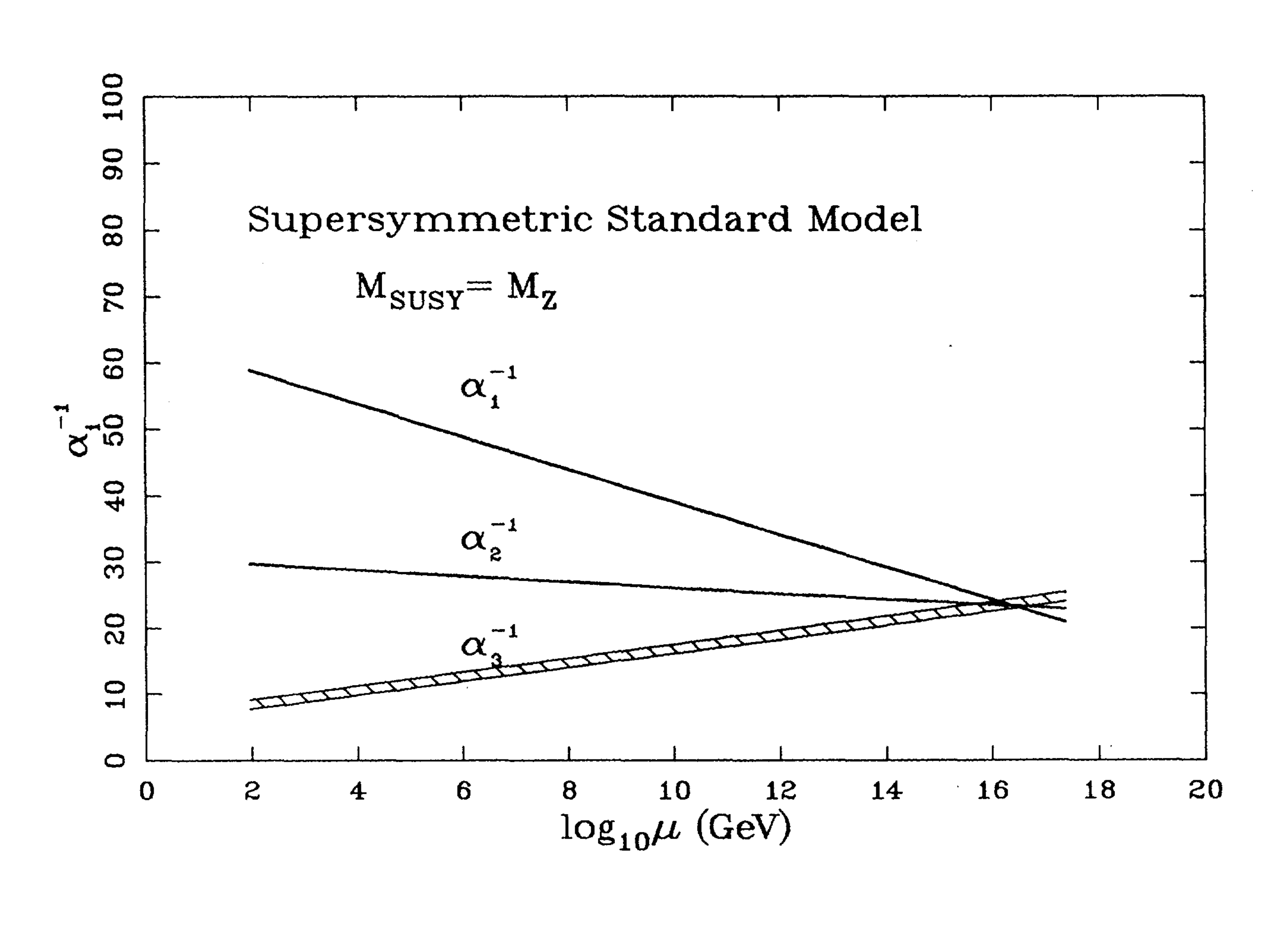}
\caption{Evolution of the three  SM gauge couplings $\alpha_i=g_i^2/(4\pi)$
as a function of $\mu=Q$ in the Supersymmetric SM \cite{Langacker:1992rq}}
\label{gumssm}
\end{figure}

The $SU(5)$ model gives us three interesting predictions: 
\begin{enumerate}
\item Hypercharge quantization.
\item Gauge coupling unification.
   \item  Proton decay.
\end{enumerate}
We already mentioned the first one.
Let us comment on the second one.
If the $SU(5)$ symmetry is exact we have that all SM gauge couplings must be equal:
\begin{equation}
g_s=g=\sqrt{\frac{5}{3}}g'\equiv g_5\, ,
\label{ggut}
\end{equation}
where the factor $\sqrt{\frac{5}{3}}$ arises from the proper normalization of $g'$.
Nevertheless, if the $SU(5)$ symmetry is broken at some scale $M_{\rm GUT}$
we only expect Eq.~(\ref{ggut}) to be fulfilled at energies above $M_{\rm GUT}$.
Indeed, in a quantum field theory the gauge couplings `runs' with  the energy  
according to the RGE. At the one-loop level we have
 \begin{equation}
\frac{d g_i}{d\ln Q}=-\frac{b_i}{8\pi^2}\,,
  \end{equation}
where $g_3=g_s$, $g_2=g$, $g_1=\sqrt{\frac{5}{3}}g'$ and $b_i$
are coefficients that depend on the spectrum of the theory.
Above $M_{\rm GUT}$ the spectrum of particles corresponds to that of a $SU(5)$
theory and we have $b_1=b_2=b_3$, but below $M_{\rm GUT}$ the $X,Y$ states and  the 
colour partner of the Higgs  are not present.  The $b_i$ are only  sensitive to  the SM spectrum; we have $b_i=(41/10,-19/6,-7)$. 
In Fig.~\ref{gusm} we plot the evolution of the three  SM gauge couplings $\alpha_i=g_i^2/(4\pi)$
as a function of $Q$.
We see that the gauge couplings tend to unify at energies around $10^{14}\ {\rm GeV}$,
  although Eq.~(\ref{ggut}) is not precisely  satisfied.
One could argue that this is a small discrepancy,   originating from  high-energy corrections to the gauge couplings.  Even so, this  implies   $M_{\rm GUT}\sim10^{14}\ {\rm GeV}$ and, as we will see later,  a  conflict with proton decay experiments.
A better situation occurs in the supersymmetric SM that we will introduce later
 motivated by the  hierarchy problem.
In this model we have   $b_i=(66/10,1,-3)$ and  a different  evolution of the gauge couplings as compared with the SM, as shown in Fig.~\ref{gumssm}.
Now the three SM gauge couplings neatly unify at energies   $\sim 10^{16}\ {\rm GeV}$,
the scale to be associated with $M_{\rm GUT}$.

Let us finally comment on proton decay.
In the $SU(5)$ model the baryon symmetry is not preserved. This is obvious
since we have put quarks and leptons in the same representation --- see Fig.~\ref{su5}.
Therefore we expect to have contributions to proton decay.
We can explicitly see that this decay is mediated by the  $X$ and $Y$ bosons
that generate the operator of Eq.~(\ref{dim6}) with $\Lambda\sim M_{\rm GUT}$.
We obtain
 \begin{equation}
\tau(p\rightarrow \pi^0e^+)
\sim 10^{34}\ {\rm years}\left(\frac{3\times 10^{15}\ {\rm GeV}}{M_{\rm GUT}}\right)^4\, .
\label{pdv}
 \end{equation}
The Super-Kamiokande detector 1000 metre underground in the Kamioka mine of 
Hida city (Gifu) Japan,
has the `titanic' task of    searching for proton decay.
This is a stainless-steel tank 39 m in diameter and 42 m tall. It is  filled with 50\ 000 tons of ultra pure water and about 13,000 photomultipliers are placed on the tank wall.
It looks for pions and positrons arising from the proton decay of the water.
Neutral pions  decay to photons that can be detected by the photomultipliers,
while positrons travelling through the water emit Cherenkov light that can also be detected
by the photomultipliers.
At present they put a bound of 
$\tau(p\rightarrow \pi^0e^+)>10^{34}\ {\rm years}$
corresponding, according to Eq.~(\ref{pdv}), to 
the bound $M_{\rm GUT}>3\times 10^{15}$ GeV.
This rules out $SU(5)$  models with  $M_{\rm GUT}\sim  10^{14}$ GeV,
and  is  at the verge of  testing
models, such as  supersymmetric $SU(5)$ models\footnote{In supersymmetric $SU(5)$ models we have other proton decay channels,  e.g., $p\rightarrow K^+\bar \nu_\tau$, that are usually more important than the one considered here \cite{Langacker:1980js}. },  where   $M_{\rm GUT}\sim  10^{16}$ GeV.

Apart from the three predictions explained above,
GUT give   other type of interesting predictions, although  they are  more model dependent.
For example in most of GUT  bottom-tau unification is predicted: $M_b=M_\tau$ at $Q\gtrsim M_{\rm GUT}$.
This prediction works  reasonable well in the supersymmetric SM. 
Nevertheless  it does not work for the other families.
Another prediction of GUT with  $G=SO(10)$ 
is the  generation of  neutrino masses through the `see-saw' mechanism.
In $SO(10)$ all SM fermions of a given family 
can be embedded in a single representation, the ${\bf 16}$ of $SO(10)$.
Apart from the SM fermions it also contains a singlet $\nu_R$ that after $SO(10)$
breaking can get  a mass
and generate the operator of Eq.~(\ref{dim5}) with $\Lambda\sim M_{\nu_R}$.
We already  saw that this operator leads to  neutrino masses of the Majorana type. This also leads to processes with 
neutrinoless double beta-decays: $nn\rightarrow ppee$. 
Thus, observing experimentally this process  is of  great importance
for understanding the nature of the neutrino masses.

\section{The strong CP-problem and axions}
\label{stroncpproblem}

In the SM Lagrangian of Eq.~(\ref{sml}) we did not include the dimension-4 operator 
$\epsilon^{\mu\nu\rho\sigma}G^a_{\mu\nu}G^a_{\rho\sigma}$ made of gluon
fields\footnote{Also the same operator but made of $W_\mu$ could be present. The impact of this term, however,  on physical observables is negligible.}.
Following the usual convention in the literature, let us introduce it as  
\begin{equation}
\frac{\theta}{32\pi^2}\int d^4x \epsilon^{\mu\nu\rho\sigma}G^a_{\mu\nu}G^a_{\rho\sigma}\, .
\label{theta}
\end{equation}
This term violates the CP symmetry
and induces  a sizeable  electric dipole moment (EDM) for the neutron. 
Experimental limits on the neutron EDM  put a limit on the coefficient $\theta$:
\begin{equation}
\theta\lesssim 10^{-10}\, .
\label{thetal}
\end{equation}
The smallness of this coefficient requires an explanation.
A possible one was proposed long ago by Peccei and Quinn \cite{Peccei:1977hh}.
They  promoted  $\theta$ to a field $a(x)$, the axion field, assumed to be a   Goldstone boson 
 arising from the  spontaneous breaking of a $U(1)$ symmetry, the PQ symmetry.
If this symmetry had a $U(1)SU(3)_c^2$-anomaly, then the only  non-derivative interaction of the axion would be  given by  the term of  Eq.~(\ref{theta})
with the replacement $\theta\rightarrow a(x)/f_A$, where $f_A$ is a dimensionful parameter called the axion decay-constant.
In this model the value of $\theta$ is dynamical and must be calculated by
minimizing the axion potential.
One obtains  $V(a)=\frac{1}{2}m^2_Aa(x)^2+\cdots$ and then,  $\langle a(x)\rangle=0\Rightarrow\theta=0$, in agreement with Eq.~(\ref{thetal}).

\begin{figure}
\centering\includegraphics[width=.95\linewidth]{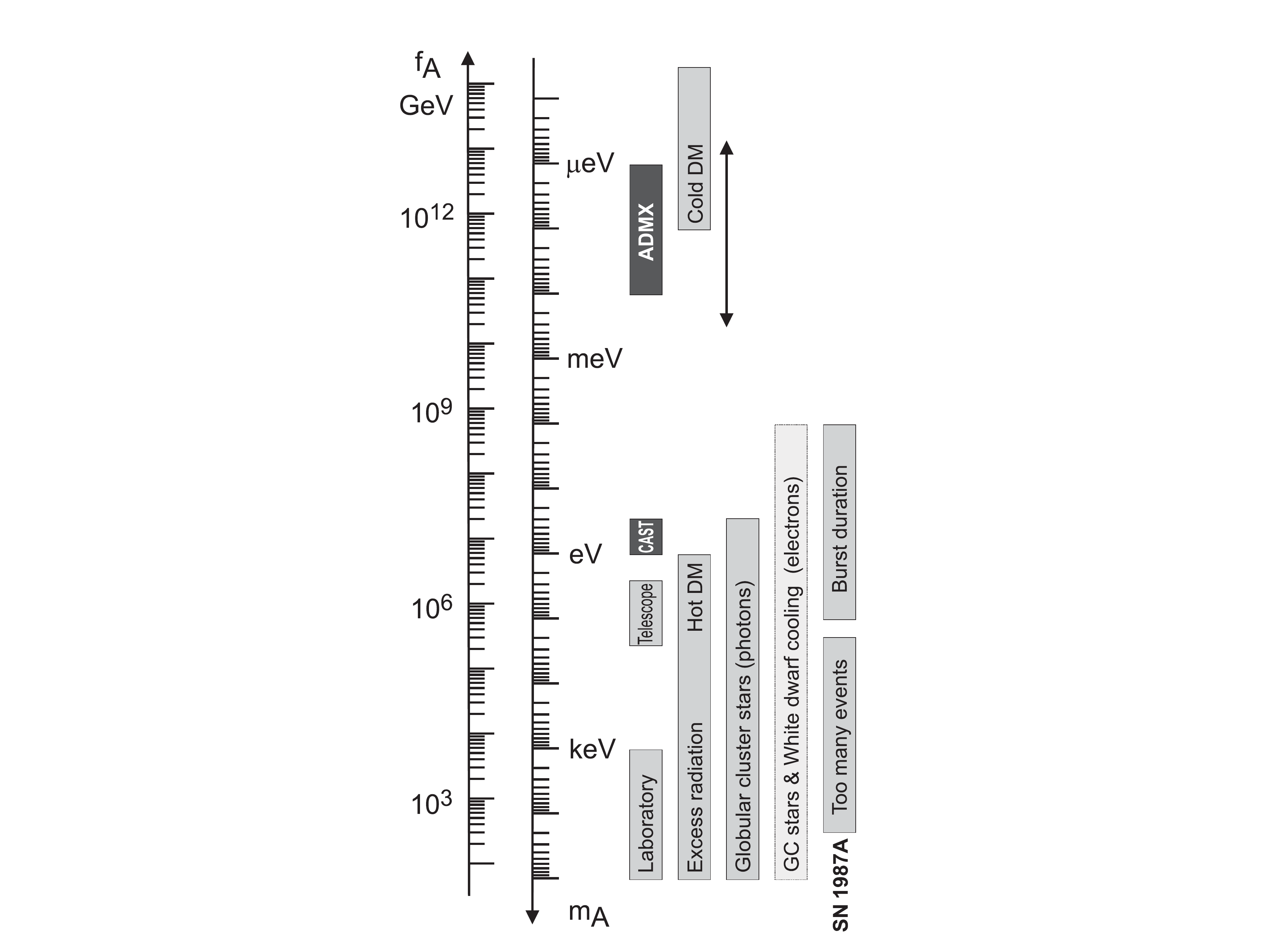}
\caption{Excluded regions of  $f_A$ (or equivalently of $m_A$)
from different experiments and astrophysical constraints. The reach of present experiments CAST and ADMX is also shown \cite{Nakamura:2010zzi}.}
\label{axionwindow}
\end{figure}

 The Peccei--Quinn mechanism  has a   testable prediction \cite{Weinberg:1977ma}.  The model  predicts the existence
 of a new particle, the axion, whose mass can be calculated. In the limit
 $f_A\gg f_\pi$, we have
 \begin{equation}
m_A=\frac{f_\pi}{f_A}\frac{\sqrt{m_u m_d}}{m_u+m_d} m_\pi\, .
\end{equation}
The axion  mass  ranges from 100 keV to $10^{-12}$ eV,
 as the unknown parameter $f_A$ varies from  100 GeV to $10^{19}$ GeV. 
Axions  couple to gluons through  Eq.~(\ref{theta})  with $\theta\rightarrow a(x)/f_A$,
so the larger $f_A$, the smaller are their couplings to SM states.
Detecting the axion would be an excellent way to prove the Peccei-Quinn idea.
Since the proposal of this mechanism, experimentalists have been searching in vain  for the axion.
Today  the values of $f_A$ (or equivalently of $m_A$) are strongly constrained,
as shown in Fig.~\ref{axionwindow}.

The axion field is a possible   dark matter candidate if $f_A$ lies around $10^{12}$ GeV.
The ADMX experiment is looking   for dark matter axions coming from the halo of the galaxy.
Since the axion couples to gluons,  it mixes with the pions, and
since these latter couple to photons, the axions also, generically,   couple to  photons.
Then  axions can scatter off a magnetic field and resonantly be converted into  microwave photons. The present searches at ADMX are aiming axions with $f_A$ between $10^{11}$~GeV and $10^{13}$~GeV as shown in Fig.~\ref{axionwindow}.

\section{The hierarchy problem}

In the SM the electroweak symmetry is triggered by the Higgs VEV.
For  positive values of $\mu^2$, the Higgs VEV is given by $v^2=\mu^2/\lambda$.
Therefore we must have $\mu^2=\lambda v^2\sim  6\times 10^{4}\lambda$ GeV$^2$.
As compared with the other dimensionful scale of the SM, the Planck scale, $M_P^2\sim 10^{38}$ GeV$^2$, the value of $\mu^2$ looks extremely small.
Why are they so different? This is the so-called hierarchy problem.

To make it worse, one can realize that the Higgs mass is, at the quantum level,
very sensitive to the mass of heavy states to which the Higgs couples.
For example, in the $SU(5)$ GUT discussed above, the Higgs couples to the 
$X$ and $Y$ bosons. 
At the one-loop, the Higgs mass squared will receive corrections  proportional to 
$M_{\rm GUT}^2/(16\pi^2)$.
Therefore it is very unnatural to expect $\mu^2\ll  M_{\rm GUT}^2$.

This sensitivity  to the heavy states is  a feature only of scalars.
For  fermions, for example,  we have that one-loop corrections to a Dirac  fermion  mass  are proportional to the fermion mass itself, and similarly for gauge bosons.
The reason for this can be easily understood using symmetries.
A Dirac fermion mass arises from the operator $m\bar\psi_L\psi_R$.
This operator is not  invariant under  the chiral phase transformation
 $\psi_R\rightarrow e^{i\theta}\psi_R$.
 This means that it can only be generated from terms in the Lagrangian
 that break this symmetry. Heavy states respecting the chiral symmetry cannot induce it.
 We say that the fermion masses are `protected' by chiral symmetries.
 Similarly for gauge bosons, the gauge symmetry protects the mass of the gauge bosons.
On the other hand, scalar  mass terms, e.g.,  $\mu^2|H|^2$, are  invariant under any phase transformation
and then can be induced by heavy virtual particles.

Three solutions have been  proposed to solve the hierarchy problem:
(1) Implement   a symmetry that  relates scalars to fermions since the masses of these latter  are not sensitive to heavy states. This is supersymmetry. (2) Assume that the Higgs is not elementary but just a composite state. (3) Assume that the only scale in particle physics is  the electroweak scale, e.g., $v\simeq 246$ GeV.
 In this case $M_P$ is not  the high-energy scale at which  gravity becomes strong,
    but just an unphysical  scale arising from  the fact that the Newton constant, that on dimensional grounds is now given by  $G_N= g_N  /v^2$, 
is extremely  small    at large distances, $g_N\ll 1$. This can be naturally
realized assuming the existence of extra dimensions.
We will discuss all of them in turn.

\section{Supersymmetry}  

Supersymmetry provides a symmetry that protects the Higgs mass.
It works in the following way. Supersymmetry relates scalars to fermions;
since the masses of these latter  are protected by (chiral) symmetries,
scalar masses will also be protected.
An  instructive  way to see this is by looking at the simplest case,
a free theory of a Majorana fermion $\Psi$ and a complex scalar $\Phi$.
Its Lagrangian is given by
\begin{equation}
{\cal L}=|\partial_\mu\Phi|^2+i\frac{1}{2}\bar\Psi\slash \hskip -.2cm \partial\Psi\, .
\end{equation}
This Lagrangian is invariant
under 
\begin{eqnarray}
\Phi\rightarrow \Phi+\delta\Phi &\qquad&  \delta\Phi=\bar\xi(1-\gamma_5)\Psi\nonumber\\
\Psi\rightarrow \Psi+\delta\Psi &\qquad& \delta\Psi=i(1-\gamma_5)\gamma^\mu\xi\partial_\mu\Phi
\, ,
\label{susy}
\end{eqnarray}
where $\xi$, the  parameter of the transformation, is a Majorana fermion (anticommuting).
Note that a mass for the scalar, $\mu^2|\Phi|^2$, is not invariant under the symmetry Eq.~(\ref{susy}). In other words, this symmetry forbids the scalar to get a mass.
Equation~(\ref{susy}) is a supersymmetry.
 It can be shown that supersymmetry is the maximal extension of the Poincar\'{e} group in a quantum field theory~\cite{Haag:1974qh}.
It contains an extra generator $Q$ that acting on fermionic states transforms them into bosonic states and vice versa\footnote{The supersymmetry considered here is often called
${\cal N}=1$ supersymmetry. There  exist extensions to this  supersymmetry  (${\cal N}=2,4$)
with a more extended algebra. They   will not be discussed here since there is no phenomenological motivation for these extensions.}.
In a schematic form the SuperPoincar\'{e} algebra is given by
\begin{eqnarray}
[Q,M_{\mu\nu}]&=&Q\, ,\nonumber\\
\{Q,Q^\dagger\}&=&P^\mu\, ,\nonumber\\
\{Q,Q\}&=&\{Q^\dagger,Q^\dagger\}=0\, ,\nonumber\\
\  [P^\mu,Q] &=&[P^\mu,Q^\dagger]=0\, .
\end{eqnarray}
The $Q$ generator computes with $P^2$ and any generator of the gauge group.
This implies that a fermion and its associated boson  have equal mass and charge.

Imposing supersymmetry on the Standard Model leads  to the Minimal Supersymmetric Standard Model
(MSSM). This is  not  a straightforward exercise, and we redirect the  interested reader 
to  Ref.~ \cite{Martin:1997ns}. Here we will only comment on the most important implications of supersymmetry
and its predictions at present and future colliders.

The most drastic implication of supersymmetry is that  the SM spectrum is required to be doubled.
For each SM quark and lepton ($Q_L$, $l_L$, ...)
 one has to add an extra  scalar,    usually called squark and  slepton ($\widetilde Q_L$, $\widetilde l_L$, ...), and  for each gauge boson and  Higgs ($W_\mu$, $H$, ...) one has to add an extra  fermion, called gauginos and Higgsinos ($\widetilde W$, $\widetilde H$, ...).
 But   this is not yet enough. This theory will have anomalies (quantum inconsistencies)
 that, to be avoided, require the addition of extra fields.  The  simplest way  is to add an extra Higgs doublet.
The Higgs sector is then given by
\begin{equation}
\begin{array}{c|c|c|c} & SU(3)_c & SU(2)_L & U(1)_Y \\\hline H_u & 1& 2 & 1 
 \\\hline H_d & 1& 2 & -1
\end{array}
\end{equation}
that are accompanied by two Higgsinos, $\widetilde H_u$
and  $\widetilde H_d$, with the same quantum numbers.
In the MSSM not all  possible terms allowed by symmetries can be added.
Some of them would lead to a violation
of the baryon and lepton symmetries, in clear contradiction with experiments.
An easy way to avoid these terms is to impose a discrete symmetry on the MSSM,
under which  all SM fields are {\it even} and all superpartners are {\it odd}.
This is an  $R$-parity.
Imposing this discrete symmetry on the MSSM  leads to  interesting consequences.
The superpartners  can only  be produced in pairs;  the lightest supersymmetric particle (LSP) is stable, resulting in a 
  dark matter candidate.
Furthermore, the  Yukawa couplings take  the form
\begin{equation}
{\cal L}_Y=Y_u^{ij}\bar Q_L^i\cdot  H_u^*\, u^j_R+
Y_d^{ij}\bar Q_L^i \cdot H_d^*\, d_R^j+
Y_e^{ij}\bar l_L^i\cdot  H_d^*\, e_R^j+h.c.
\end{equation}
The Yukawa couplings then become very sensitive to  the ratio of the 
Higgs VEVs, $\langle H_u\rangle/\langle H_d\rangle\equiv \tan\beta$,
since we have
\begin{equation}
 Y_{u}=\frac{ g M_{u} }{\sqrt{2} M_W \sin\beta}\ , \ \ \ 
 Y_{d,e}=\frac{ gM_{d,e}} {\sqrt{2} M_W \cos\beta}\, .
\end{equation}

\begin{figure}
\centering\includegraphics[width=.8\linewidth]{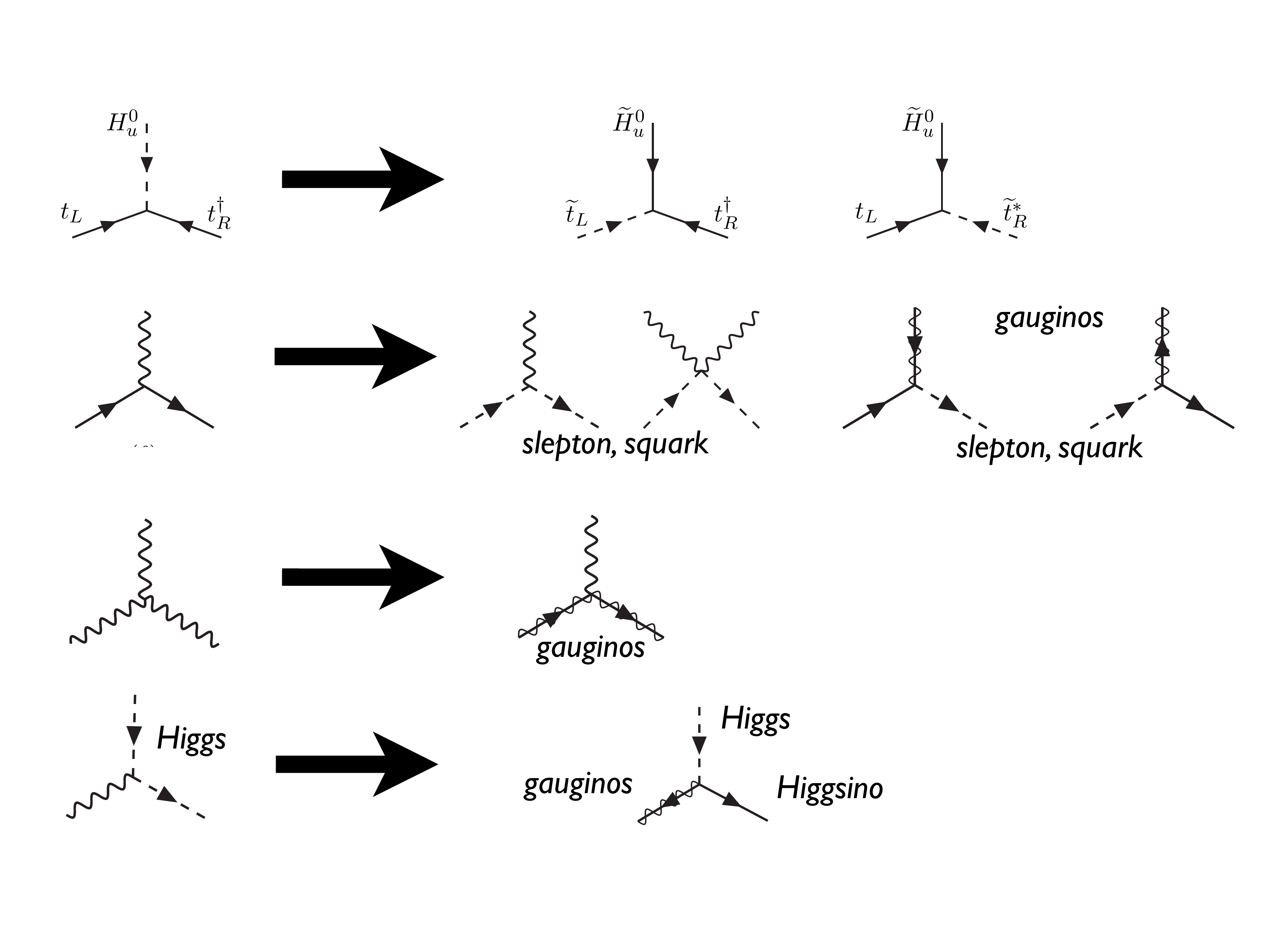}
\caption{MSSM interactions obtained from `supersymmetrization' of the SM interactions}
\label{mssmc}
\end{figure}

Although the derivation of all  MSSM interactions is a difficult task, it is relatively easy 
to obtain the main interactions needed for phenomenology. 
They can be obtained  by `supersymmetrization' that corresponds to taking any SM 
interaction and replacing $fermion\leftrightarrow boson$ consistently with 
the SM symmetries. This is depicted in Fig.~\ref{mssmc}.
The only interactions not obtained in this way are scalar trilinears  and quartics.

\begin{figure}
\centering\includegraphics[width=.6\linewidth]{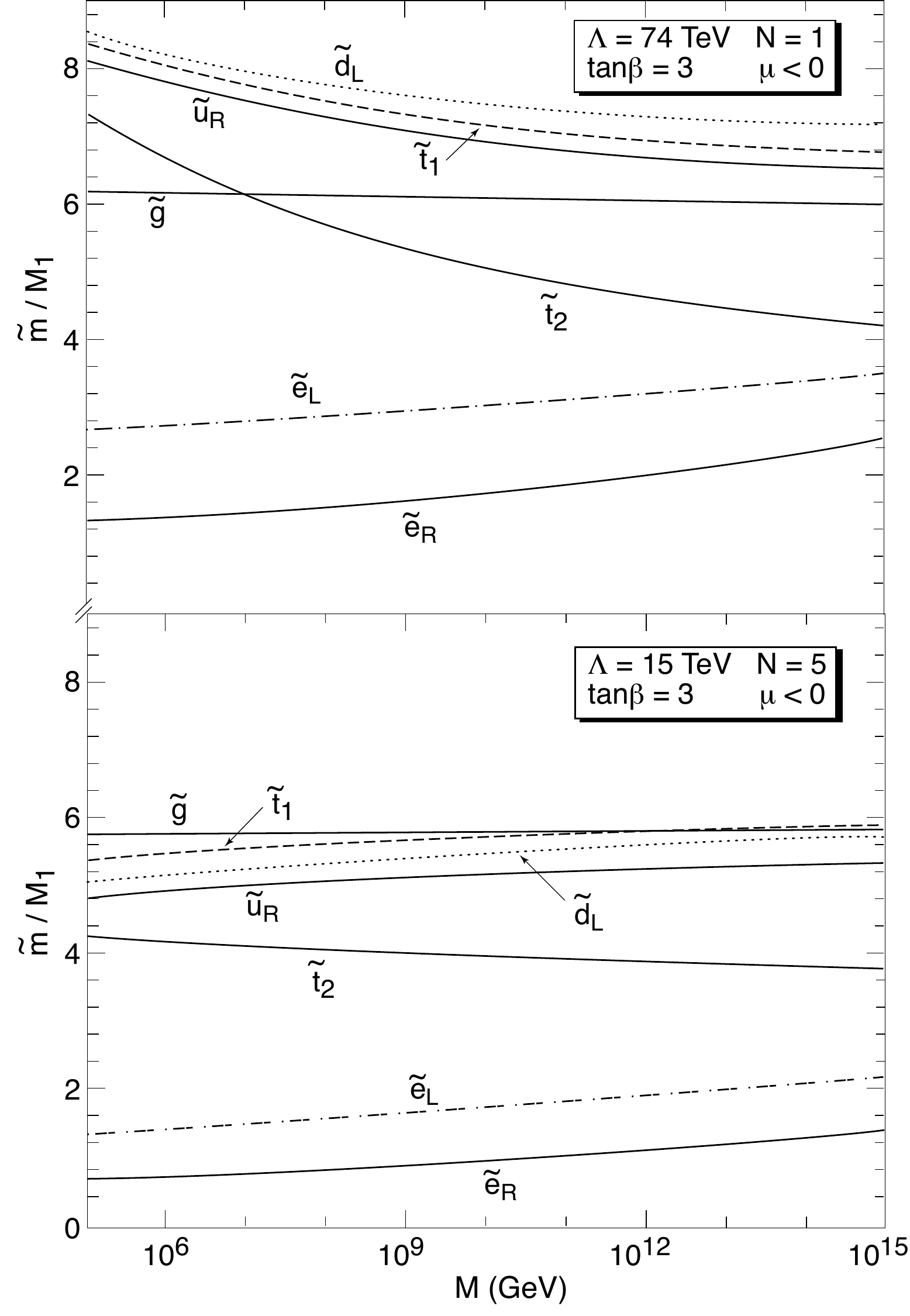}
\caption{Superparticle mass spectrum, normalized to the bino mass $M_1$, in GMSB models taken from Ref.~\cite{Giudice:1998bp}. Here  
$\Lambda=F/M$ and $N$ is the number of messengers.}
\label{gmsbs}
\end{figure}

If supersymmetry is exact, the mass of a fermion and that of its associated boson must be 
the same. This implies that, for example, s-electrons must have mass of half MeV.
We have not seen such a light state, implying that supersymmetry must be broken.
Like in the SM with the electroweak symmetry, 
we can assume that supersymmetry is  spontaneously broken such that 
all superpartners  receive mass.  These masses cannot be much larger than the electroweak scale,
otherwise we will have back the hierarchy problem that was the main motivation for supersymmetry. Indeed, after supersymmetry breaking,  corrections to the Higgs masses
become  proportional to the superpartner masses; these latter must then be  kept around the electroweak scale.
Several realistic  models of supersymmetry breaking have been proposed in the literature.
The simplest one is called Gauge Mediated Supersymmetry Breaking (GMSB) \cite{Giudice:1998bp}.
It requires an extra  sector responsible for spontaneous breaking of supersymmetry, 
containing    fields  charged under the SM gauge group, the `messengers'.
The MSSM only knows about supersymmetry breaking due to the gauge interactions;
 therefore  gaugino, squark, and slepton  masses  arise at the loop level 
from   these interactions that mediate the breaking from the supersymmetry-breaking sector
to the MSSM sector.
The model is quite predictive. Up to EWSB effects,
gaugino and scalar masses depend on only two parameters: the supersymmetric breaking scale $\sqrt{F}$ and the mass of the messengers $M$.
The scalar masses are the same for all families, guaranteeing the absence of 
dangerous flavour-violating  interactions.
This is a crucial requirement to obtain  realistic scenarios of supersymmetry breaking. 
The  generation of  the Higgsino mass is a difficult task
in GMSB models   and requires an extension of the model \cite{Dvali:1996cu}.
After EWSB, the spectrum is (slightly) modified and becomes sensitive to $\tan\beta$.
A typical spectrum is shown in Fig.~\ref{gmsbs}. The colour states are the heaviest, while
right-handed sleptons, that have only hypercharge interactions,  and binos $\widetilde B$,
are the lightest.
The LSP, however, not shown in Fig.~\ref{gmsbs}, is the superpartner of the graviton, the gravitino. Its mass is given by
\begin{equation}
m_{3/2}=\frac{F}{k\sqrt{24\pi}M_P}=\frac{1}{k}\left( \frac{\sqrt{F}}{100~
{\rm TeV}}\right)^2 2.4~{\rm eV}~,
\label{gravmass}
\end{equation}
where $k$ is a model-dependent coefficient  such that
$k<1$.

Another  popular scenario of supersymmetry breaking is the so-called minimal  supergravity
model or constrained MSSM \cite{Chamseddine:1982jx}. 
It is usually presented as a model that predicts,
at  energies $Q\simeq M_{\rm GUT}\sim 10^{16}$ GeV,
universal  gaugino masses, $M_{1/2}$,  universal scalar masses, $M_0$,
and universal  trilinears, $A_0$.
These `predictions' however are not justified by any symmetry.
Therefore one  must consider this scenario just as a simplified Ansatz on the MSSM parameters, 
and not as a model. 
The LSP in this scenario is  the neutralino, $\tilde \chi^0$, that is a mixture of neutral gauginos and Higgsinos.

\begin{figure}
\centering\includegraphics[width=.90\linewidth]{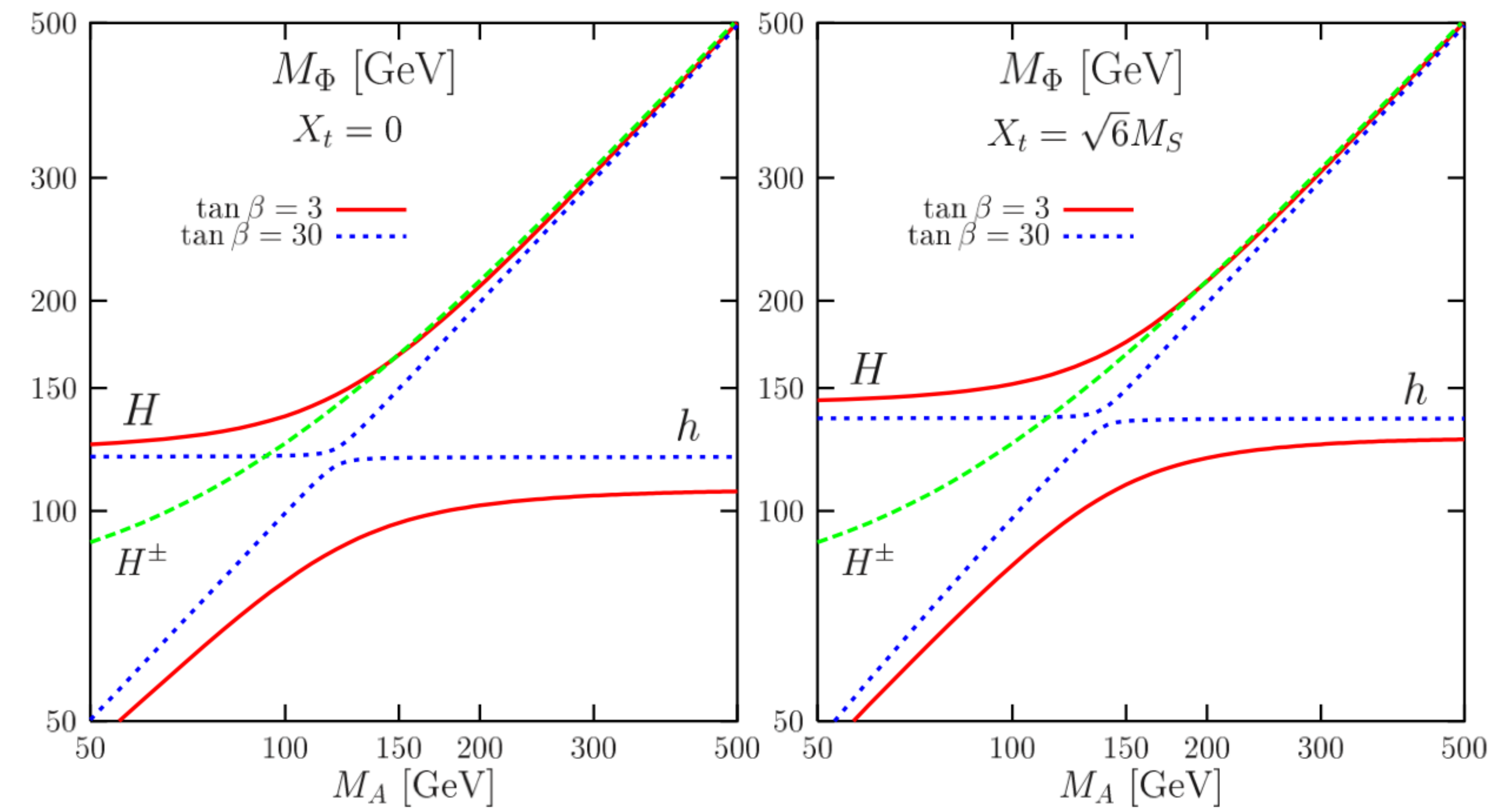}
\caption{Higgs spectrum in the MSSM  taken from Ref.~\cite{Djouadi:2005gj}.}
\label{higgsmssm}
\end{figure}

\begin{figure}
\centering\includegraphics[width=.90\linewidth]{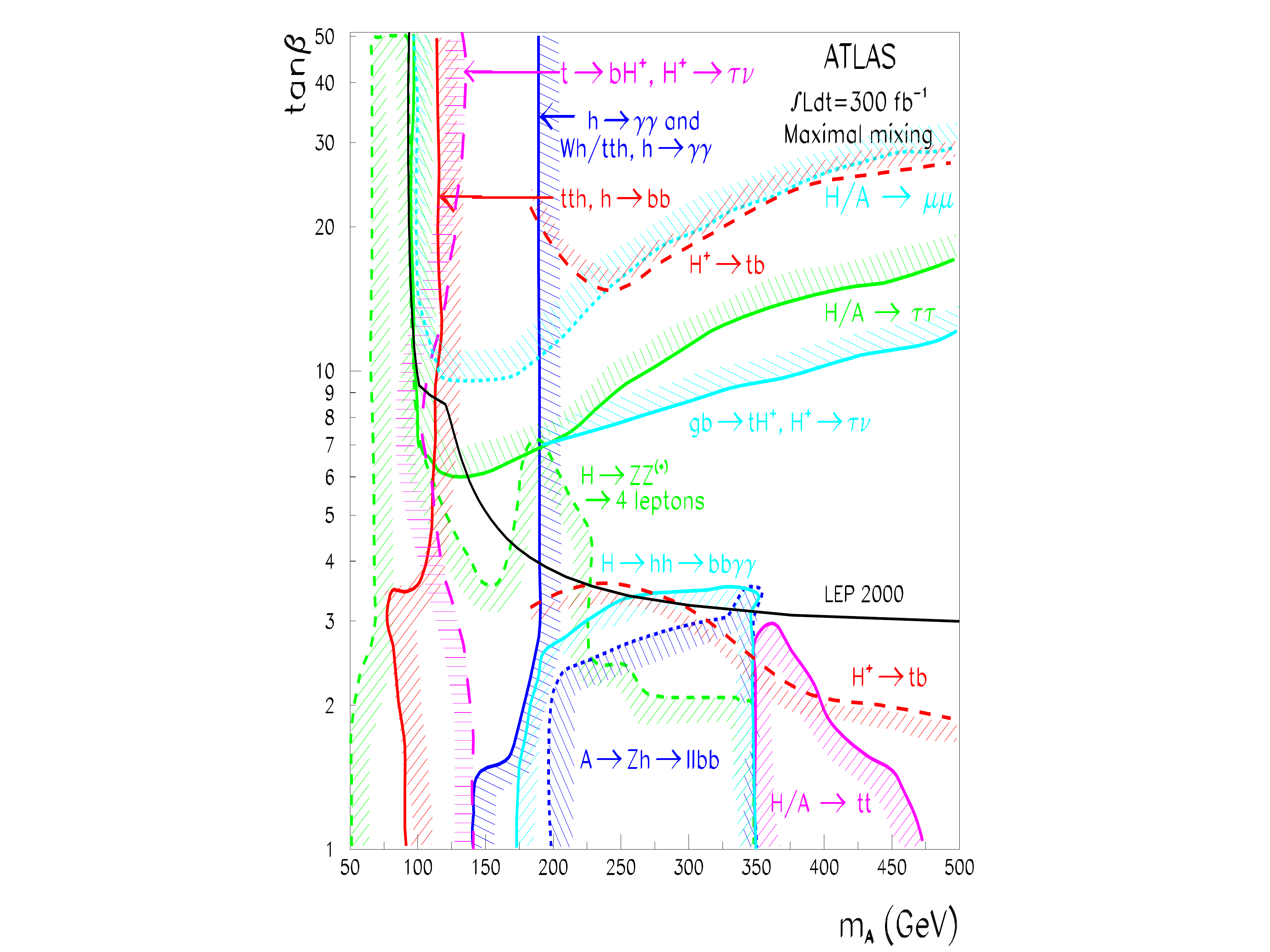}
\caption{Decay channels of the MSSM Higgs  bosons that will be  
 visible at the LHC as a function of  $\tan\beta$ and $M_A$}
\label{higgssearch}
\end{figure}

A quite constrained sector of the MSSM is the  Higgs sector.
The Higgs potential is given by 
\begin{eqnarray}
V(H_u,H_d)&=&m_1^2 |H_u|^2+
m_2^2|H_d|^2+m_{12}^2\, H_u\cdot H_d+h.c.\nonumber\\
&+&\frac{g^2+g'^2}{8}(|H_u|^2-|H_d|^2)^2+
\frac{g^2}{2}|H_u\cdot H_d|^2\, .
\end{eqnarray}
The spectrum consists of five physical Higgs bosons, two CP-even neutral, $h^0$ and $H^0$, one CP-odd, $A^0$ and two charged, $H^\pm$.
The Higgs potential
 depends only on three unknown parameters. One  is fixed by $v^2=2(\langle H_u\rangle^2+\langle H_d\rangle^2)$, while  the other two can be traded by $\tan\beta$ and $M_A$.
At tree level, the other Higgs masses are given by
\begin{eqnarray}
M^2_{H^+}&=&M^2_A+M^2_W\nonumber\, ,\\
M^2_{h,H}&=&\frac{1}{2}\left\{M^2_A+M^2_Z\mp \sqrt{(M^2_A-M^2_Z)^2+4\sin^2 2\beta \, M^2_AM^2_Z}\right\}\, ,
\end{eqnarray}
that leads to the prediction $M_h\leq M_Z$.
Quantum corrections however  change this prediction, making $M_h$  very sensitive to
the superparticle spectrum. In spite of this, the Higgs $h^0$ is always light, $M_h\lesssim 130$~GeV (see Fig.~\ref{higgsmssm}) and should be visible at the LHC through its decay to $bb$, $\tau\tau$ or $\gamma\gamma$.
The other Higgs bosons will also be visible at the LHC (Fig.~\ref{higgssearch}) except in certain regions of $\tan\beta$--$M_A$ where they are difficult to be seen.

\subsection{Superpartners at Hadron colliders}

The hunting of superparticles at  the Tevatron and the LHC
is quite involved  due to the large number of particles.
In models with   $R$-parity the superpartners are always produced in pairs and cascade 
down to the LSP that, being stable, goes away from the detectors.
A typical  example is  gaugino hunting. Gauginos,
once produced, can  decay through different channels.
For example, if $ \tilde \chi^0$ is the LSP,  we can have
 \begin{equation}
 \tilde g\rightarrow q\tilde q\rightarrow qq\tilde\chi^0\, .
 \end{equation}
 In this case the signal consists of  looking for an excess of  jets+ missing $E_T$.
In certain cases the decay can be 
 \begin{equation}
\tilde g\rightarrow q\tilde q\rightarrow  qq\tilde \chi^+\rightarrow qql^+\tilde \nu
\rightarrow qql^+\nu\tilde\chi^0\, ,
 \end{equation}
 where $\tilde \chi^+$ is a chargino, a mixture of $\widetilde W$ and charged Higgsinos.
 The final  signal has in this case two extra leptons that are easy to detect.
For  GMSB models, where the LSP is the gravitino, we have instead
that $\tilde\chi^0$ decays to the gravitino,  $\tilde \chi^0\rightarrow \gamma+\tilde G$,
and the final signal can be accompanied by  photons.
Searches on  charginos  require similar  signatures, although the production
cross-sections are obviously smaller.

\begin{figure}
\centering\includegraphics[width=.5\linewidth]{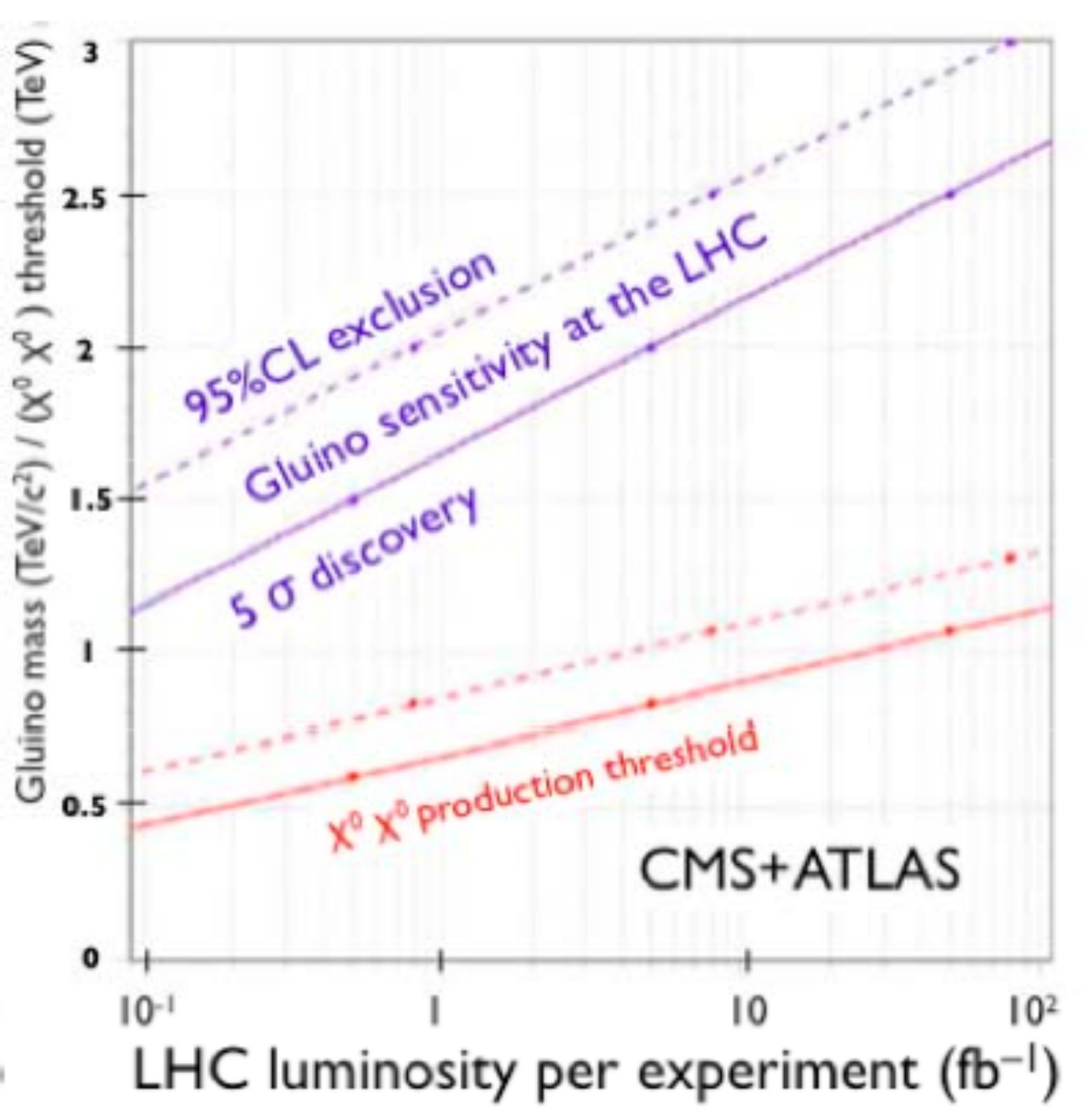}
\caption{Expected sensitivity  at the LHC for $\tilde g$  and  $\tilde \chi^0$ searches  for different  luminosities}
\label{mssmreach}
\end{figure}

In almost all cases,  the Tevatron and, especially,  the LHC can do a good job and
reach  superparticles  up to very high masses.  
Figure~\ref{mssmreach} shows the expected sensitivity  at the LHC for gaugino and neutralino searches  for different  luminosities. If supersymmetry is there, we have 
a good chance to discover it.

\section{Higgsless and composite Higgs}
\label{strongsec}

Although the Higgs mechanism  is a simple and economical
way to break the electroweak gauge symmetry of the SM
and at the same cure the bad high-energy behaviour of the  $W_LW_L$ scattering amplitudes,
it has, as we showed, an `expensive  price to pay': the hierarchy problem.
For this reason,  it is interesting to  look for other ways to break the electroweak symmetry and unitarize the $W_LW_L$ scattering amplitudes.
An example can be found in QCD, where pion-pion scattering is unitarized 
by the additional resonances that arise from the SU(3)$_c$  strong dynamics.
A replica of QCD at  energies $\sim$ TeV that breaks the electroweak symmetry
can then be  an alternative  to the Higgs mechanism. 
This is the so-called  technicolor model~\cite{TC} (TC).  
In TC there is no  Higgs particle and the SM scattering amplitudes 
are unitarized, as in QCD,  by infinite heavy resonances.  
One of the main obstacles to implementing  this approach 
has arisen from EWPT 
that has disfavoured this type of model.
The reason has been the following. 
Without a Higgs, we expect the new particles responsible for unitarizing 
the SM amplitudes to have a mass at around 1 TeV.
These same resonances give large tree-level contributions
to the electroweak observables.

There have been two  different alternatives to  overcome  this problem.
Either one
assumes that (1) there are  extra contributions to the electroweak observables 
that make the model  consistent with the experimental data, or
 (2) that the strong sector does not break the electroweak symmetry but it
just delivers a composite pseudo-Goldstone boson (PGB)  to be identified with  the Higgs.
This Higgs gets a potential at the one-loop level and triggers EWSB 
at lower energies.

In the first case, the Higgsless approach, EWPT are satisfied
thanks to additional   contributions to the electroweak observables that can come
 from extra scalars or fermions
of the  TC model,
or from vertex corrections.
As we will see, the cancellations  needed to pass the EWPT
are not  large, making this possibility not so inconceivable.

In the second case, the Higgs 
 plays the  role of  
 {\it partly} unitarizing  the SM scattering amplitudes.
Compared to theories without a Higgs, 
the scale at which new dynamics is needed can be  delayed, and therefore
the extra resonances that ultimately unitarize the SM amplitudes
can be heavier.
In this case  EWPT will be under control.
This is the approach of the composite Higgs models, first considered 
by Georgi and Kaplan~\cite{GK}.
In these theories a light Higgs arises as a PGB
of a strongly interacting theory, in a very similar way to pions in QCD.

Although these scenarios offer an interesting completion of 
 the SM, 
the difficulty of calculating within strongly coupled theories 
 has been a deterrent from fully exploring them.  
Nevertheless, the situation has changed  in recent years.
Inspired by the AdS/CFT correspondence \cite{adscft}, 
a  new approach to building realistic and predictive  Higgsless and 
  composite Higgs  models  has been developed. 
The AdS/CFT correspondence
 states that weakly coupled five-dimensional  (5D) theories in   Anti-de-Sitter (AdS)  have a 4D holographic description 
in terms of strongly coupled conformal field theories (CFT).
Such correspondence gives a definite prescription on how to construct five-dimensional theories 
that have the same physical behaviour and symmetries as the desired strongly coupled 4D theory.
This has allowed one to propose  concrete Higgsless \cite{Csaki:2003zu}
 and  composite Higgs \cite{acp,cdp} models 
that not only are consistent with the experimental constraints, but also 
give clear predictions  for  the physics at the LHC. 
We will briefly discuss them in the section on extra dimensions.

\subsection{The original Technicolor model. Achievements and pitfalls}
Technicolor models \cite{TC} of EWSB consist of  a new strong gauge sector, 
$SU(N)$ or $SO(N)$,  that it is assumed
to  confine at a low-scale $\mu_{IR}\sim$ TeV.
In addition, the model contains (at least) two flavours of 
techni-quarks  $T^{u,d}_L$, $T^{u,d}_R$ transforming
in the fundamental representation of the strong group and 
as ordinary quarks under the  electroweak group.
As occurs in QCD, this implies that the strong sector has 
a global  $G=SU(2)_L\times SU(2) _R\times  U(1) _X$ symmetry
under which $T^{u,d}_{L}$  transforms as a $\bf (2,1)_{1/6}$
and $T^{u,d}_R$  transforms as a $\bf (1,2)_{1/6}$
(the hypercharge is  given by $Y=2(T_3^R+X)$).
Assuming  that the TC quarks condensate,
$\langle\bar T_LT_R \rangle\sim\mu^3_{IR}$,
the global symmetry of the strong sector 
$G$ is broken down to  $H=SU(2)_V\times U(1)_X$.
The electroweak symmetry is then broken giving masses to the corresponding SM gauge bosons. Fermion masses are assumed to arise from higher-dimensional operators 
such as 
$\bar q_L u_R\bar T_RT_L/M^2$ that can be induced  from an extended heavy gauge 
sector (ETC).
After the TC-quark condensation, SM fermions  get masses $m_u\sim \mu^3_{IR}/M^2$.

If the number of colours $N$ of the TC group is large enough, the strong sector  
can be described by an infinite 
number of resonances \cite{'thooft}. 
The masses and couplings  of the resonances
depend on  the model. Nevertheless,  as in QCD,
 we can expect vector resonances
transforming as a triplet of $SU(2)_V$; the TC-rho of mass
$m_\rho\sim \mu_{IR}$.
In order to see the implications of these resonances on 
the SM observables, it is useful to write the low-energy Lagrangian
of the SM fields obtained 
after  integrating out the strong sector (the equivalent of the QCD chiral Lagrangian).
It is convenient to express
this Lagrangian in a $SU(2)_L\times SU(2)_R\times U(1)_X$-symmetric way.
To do so, we promote the elementary SM fields  to
fill complete  representations of $SU(2)_L\times SU(2)_R\times U(1)_X$.
For the bosonic sector, this means to  introduce extra non-dynamical vectors, {i.e.}, spurions,  
to complete the corresponding adjoint  representations
 $W_\mu^L$,  $W_\mu^R$, and  $B_\mu$.
Having the Goldstone multiplet $U$ parametrizing  the coset 
$SU(2)_L\times SU(2)_R/SU(2)_V$,
the bosonic low-energy Lagrangian is given by 
\begin{equation}
{\cal L_{\rm eff}}=f^2\Big[\frac{1}{4}|D_\mu U|^2+
\frac{ c_S}{m^2_\rho}{\rm Tr}[W^L_{\mu\nu}UW^{R\, \mu\nu}U^\dagger]+\cdots\Big]\, ,
\label{effectivel}
\end{equation}
where
$D_\mu U=\partial_\mu U+iW^L_\mu U-iU W^R_\mu $ and
$f$ is the analog of the pion decay constant that scales as
$f\sim \sqrt{N}/(4\pi)\times m_\rho$ \cite{'thooft}. 
In Eq.~(\ref{effectivel}) we have omitted terms of order 
$(DU)^4$ that do not contribute to the SM gauge boson self-energies, and
terms of order $f^2 D^2/ m_\rho^4$ 
that are subleading for physics at energies below $m_\rho$.
The  $c_S$ is an order-one coefficient that in QCD takes the value
$c_S=L_{10} m^2_\rho/f^2\simeq -0.4$. 
The mass of the SM $W$ arises from the kinetic term of $U$
that  gives   $M_W^2=g^2f^2/4$
from which we can deduce
\begin{equation}
 f= v\simeq 246\ {\rm GeV}\ \ \ \ {\rm and}\ \ \ \  m_\rho\simeq 2\, \sqrt{\frac{3}{N}}\ {\rm TeV}\, .
\end{equation}
We also obtain Eq.~(\ref{rhop}) due to the $SU(2)_V$ symmetry that corresponds to  a custodial symmetry.

\subsubsection{Flavour-changing neutral current (FCNC) and the top mass}

If the SM fermion masses
arise from an ETC sector 
that generate the operators
$\bar q^i_Lu^j_R\bar T_RT_L/M^2$,
this sector also will generate  FCNCs  of order 
$\bar q^i_Lu^j_R\bar q^k_Lu^l_R/M^2$ that are  larger than experimentally allowed.
Also the top mass is too large to be generated from a higher dimensional
operator. 
Solutions to these problems have been  proposed \cite{review}. 
Nevertheless most of the solutions cannot successfully pass EWPT.

\subsubsection{Electroweak precision tests}

The most important  
corrections to the electroweak observables  coming from 
TC-like models are universal corrections to the SM gauge boson
self-energies, $\Pi_{ij}(p)$, and  non-universal corrections to $Zb\bar b$,
$\delta g_b/g_b$.
The universal corrections to the SM gauge bosons can
be parametrized by four quantities: 
$\widehat S$, $\widehat T$, $W$ and $Y$ \cite{Barbieri:2004qk}.
The first two,  the most  relevant ones for TC models \cite{PT}, 
are defined as  
\begin{equation}
\widehat S=g^2\Pi^\prime_{W_3B}(0)\ ,\quad 
\widehat T=\frac{g^2}{M^2_W}\left[\Pi_{W_3W_3}(0)-
\Pi_{W^+W^-}(0)\right]\, .
\end{equation}
Since $\widehat T$ is protected by the custodial symmetry,  the Lagrangian Eq.~(\ref{effectivel}) only generates $\widehat S$. We have 
\begin{equation}
\widehat S=-g^2c_S\frac{f^2}{m^2_\rho}\simeq  2.3\cdot 10^{-3}\left(\frac{N}{3}\right)\, ,
\label{scont}
\end{equation}
where we have extracted the result from QCD.
Extra contributions to  $\widehat S$, beyond those of the SM,
are  constrained by the experimental data.
They must be smaller than\footnote{Since TC models
do not have a Higgs, we are taking the result of Ref.~\cite{Barbieri:2004qk}
for $M_h\simeq 1$ TeV.}
 $\widehat S\lesssim 2\cdot 10^{-3}$ at  99$\%$ CL.
We   see that the contribution Eq.~(\ref{scont}) is at the edge of the allowed value.
Models with $N$ larger than 3 or with an extra generation of   TC quarks,
needed for realistic constructions  (ETC models),
are therefore ruled out.
The bound  $\widehat S\lesssim 2\cdot 10^{-3}$
can be  saturated  only if   $\widehat T$ 
receives extra positive contributions $\sim 5\cdot 10^{-3}$ beyond those of the SM.
 Although
 the custodial $SU(2)_V$ symmetry of the TC models
guarantees the vanishing of the TC contributions to the $\widehat T$-parameter,
 one-loop contributions involving  both the top and the TC sector are nonzero.
Nevertheless, in strongly interacting theories  we cannot reliably calculate 
 these contributions and know whether  they give the right amount
to $\widehat T$.

As we said before, the generation of a  top mass around the experimental value  
is difficult to achieve in TC models and
  requires new strong dynamics beyond
the original sector \cite{review}.
Even if a large enough top mass is generated,  an extra difficulty arises from
$Zb\bar b$.
On dimensional grounds, assuming that $t_{L,R}$ couples with equal strength
to the TC sector responsible for EWSB, we have the estimate
$\delta g_b/g_b\sim m_t/m_\rho\gtrsim 0.07$
that overwhelms the 
experimental bound $|\delta g_b/g_b|\lesssim 5\cdot 10^{-3}$.
Similar conclusions are reached even if
$t_{L,R}$ couples with different  strength
to the TC sector \cite{acp}, unless the custodial symmetry is preserved by  the  $b_L$ coupling\cite{Agashe:2006at}.

Realistic extra-dimensional Higgsless models  can be constructed  in which
the above problems can be overcome, although  this  requires extra new assumptions
and some adjustments of the parameters of the model  \cite{Cacciapaglia:2006gp}.

\subsection{Composite PGB  Higgs}
\label{ch}

By enlarging the  group $G$, while keeping qualitatively the same properties
of the Higgsless models described above, we are driven to a  
different scenario in which the strong sector 
instead of breaking the electroweak symmetry,
contains  a light Higgs  in its spectrum that will be the  responsible for EWSB.  
The  minimal model consists of a strong sector with the following symmetry breaking pattern~\cite{acp}:
\begin{equation}
SO(5) \rightarrow SO(4)\, .
\end{equation}
It  contains four  Goldstone bosons   parametrized by  the $SO(5)/SO(4)$  coset:
\begin{equation}
\Sigma = \langle \Sigma\rangle e^{\Pi/f} \ ,
 \qquad \langle \Sigma\rangle =  (0,0,0,0,1) \ ,
 \qquad 
 \Pi = 
\left(
\begin{array}{cc}
0_{4} & h_a \\
- h^T_a & 0\\
\end{array}
\right)
\, ,
\end{equation}
where $h_a\  (a=1,...,4)$ is  a  real 4-component vector, which transforms as a doublet 
under  $SU(2)_L\in SO(4)$.
This is identified with the Higgs.
Instead of following the TC idea for fermion masses described before, we can 
assume, inspired by extra dimensional models \cite{acp},
that the SM fermion couples linearly to fermionic resonances of the strong sector.
This can lead to  correct fermion masses without severe FCNC problems.

The low-energy theory for the PGB Higgs, written in a SO(5)-invariant way,
is  given by
\begin{equation}
{\cal L}_\text{\rm eff}=f^2\left[
 \frac{1}{2} \left(D_\mu \Sigma\right) \left(D^\mu \Sigma\right)^T
  + \frac{c_S}{m^2_\rho} \Sigma 
F_{\mu\nu}
F^{\mu\nu} 
\Sigma^T+V(\Sigma) +
\dots \right]\, ,
\label{effmchm}
\end{equation}
where $F_{\mu\nu}$ is the field-strength of the  $SO(5)$ gauge bosons (only the SM bosons must be considered dynamical).
From the kinetic term of $\Sigma$ we obtain
$M_W^2 = g^2 (s_h\,  f)^2/4$ together with Eq.~(\ref{rhop}), where we have defined $s_h\equiv\sin h/f$ with $h=\sqrt{h^2_a}$.
This  implies
\begin{equation} \label{eq:vbsm}
v =s_h f \simeq 246\ {\rm GeV} \, .
\end{equation}
In this model the contribution to $\widehat S$    has an extra suppression factor $v^2/f^2$ as compared to  Eq.~(\ref{scont}),
and then for $v\ll f$ one can 
satisfy the experimental constraint. 
Also $\delta g_b/g_b$ can be under control due to the custodial symmetry~\cite{Agashe:2006at}.
The exact value of $v/f$ comes from minimizing the Higgs potential $V(h)$
that arises at the loop level from
 SM couplings to the strong sector that break the global $SO(5)$ symmetry.
 The dominant
contribution comes at one-loop level from the elementary $SU(2)_L$ gauge
bosons and top quark.
In the model of Ref.~\cite{cdp},
the potential is approximately given by
\begin{equation}
V(h)\simeq \alpha\  s^2_h-\beta\  s_h^2 c_h^2\, ,
\label{potential}
\end{equation} 
where $\alpha$ and $\beta$ are constants induced at the one-loop level.
For $\alpha< \beta$ and  $\beta\geqslant 0$ we have that the electroweak symmetry is broken
and, if $\beta>|\alpha|$, the minimum of the potential is at
\begin{equation}
s_h=\sqrt{\frac{\beta-\alpha}{2\beta}}\, .
\label{vev}
\end{equation}
To have $s_h< 1$ as required, we need   $\alpha\sim \beta$ that can be accomplished in certain regions of the parameter space of the models.
The physical Higgs mass is given by 
\begin{equation} \label{mHiggs}
M_{h}^2 \simeq\frac{8\beta s^2_hc^2_h}{f^2}\, .
\end{equation}
Since $\beta$ arises from one-loop effects, the Higgs  is light.
In  the extra-dimensional composite Higgs models  \cite{cdp}
one obtains $f\gtrsim 500$~GeV,  $m_\rho\gtrsim 2.5$~TeV, and
$M_h\sim 100$--$200$~GeV.

In recent years similar  ideas based on  Higgs as a PGB 
have also been put forward under the name of 
 Little Higgs (LH) models \cite{ArkaniHamed:2001nc}.  
In these models, however, the gauge and fermion sector is extended in order
to guarantee that  Higgs  mass corrections arise at the two-loop level instead of one-loop,
allowing for a better insensitivity of the electroweak scale  to the strong sector scale $m_\rho$.

\subsection{LHC phenomenology}
\label{lhc}
\subsubsection{Heavy resonances at the LHC}

The universal feature of strongly  coupled theories of EWSB 
or their extra dimensional analogs
is   the presence of  
vector  resonances,   triplet under $SU(2)_V$, of masses in the range 
 $0.5$--$2.5$~TeV; they are the TC-rho or Kaluza--Klein states  of the $W_\mu$. 
They can either  be
produced in a $q\bar q$ Drell--Yan scattering or 
via weak boson fusion.
These vector resonances  will mostly decay into pairs of longitudinally polarized weak bosons
(or, if possible, to a weak boson plus a Higgs), and to pairs of tops and bottoms.
Studies at the LHC have been devoted to a very light TC-rho, $m_\rho\lesssim 600$ GeV,
that will be able to be seen for  an integrated luminosity of $4$ fb$^{-1}$ \cite{Ball:2007zza}.

In extra-dimensional Higgsless and composite Higgs models 
one also expects heavy gluon resonances.
Their dominant production mechanism at the LHC is 
through $u\bar u$ or $d\bar d$ annihilation, decaying  
 mostly in  top pairs. The signal will then be a 
  bump in the invariant $t\bar t$ mass distribution.
For an integrated luminosity of  $100$ fb$^{-1}$ the reach of the gluon resonances
can be up to masses of 4 TeV \cite{Agashe:2006hk}.

The most promising way to unravel   
some composite Higgs model is by
detecting  heavy fermions  with electric charge $5/3$ ($q^*_{5/3}$) \cite{cdp}.
For not-too-large values of its mass $m_{q^*_{5/3}}$, roughly below 1 TeV, these new particles will  be mostly produced in pairs, via QCD interactions,
\begin{equation}
q\bar q , gg \to q^*_{5/3}\, \bar q^*_{5/3} \, ,
\label{pro}
\end{equation}
with a cross-section completely determined in terms of $m_{q^*_{5/3}}$. 
Once produced, $q^*_{5/3}$  will mostly decay  to a (longitudinally polarized)
$W^+$ plus a top quark.
The final state of the process Eq.~(\ref{pro}) 
 consists then  mostly  of four $W$'s and two $b$-jets:
\begin{equation}
q^*_{5/3}\, \bar q^*_{5/3} \to W^+ t \, W^- \bar t \to W^+ W^+ b \, W^- W^- \bar b \, .
\end{equation}
Using same-sign dilepton final states
we could discover these particles for masses of 500 GeV (1 TeV)
for an integrated luminosity of 100 pb$^{-1}$ (20 fb$^{-1}$) \cite{Contino:2008hi}.
For increasing   values of $m_{q^*_{5/3}}$ the cross-section for pair production quickly drops,
and single production might become more important; masses up to 1.5~TeV could be reached at the LHC  \cite{Mrazek:2009yu}.

Besides  $q^*_{5/3}$, certain composite models and LH models also predict  states of electric charge $2/3$ or 
$-1/3$ that could  also be   produced in pairs via QCD interactions 
or singly via $bW$ or $tW$ fusion~\cite{Han:2003wu,Azuelos:2004dm}.
They will decay to a SM top or bottom quark plus a longitudinally polarized $W$ or
$Z$, or a Higgs.
When kinematically allowed, a heavier resonance will also decay to a lighter one accompanied  
with  a  $W_{\rm long}$,  $Z_{\rm long}$ or $h$.
Decay chains could lead to extremely characteristic final states. For example, in 
one of the models of Ref.~\cite{cdp}, 
the Kaluza--Klein with charge $2/3$  is predicted
to be generally heavier than $q^*_{5/3}$.
If pair produced, they  can decay to $q^*_{5/3}$ leading to a spectacular six $W$'s plus 
two $b$-jets final state:
\begin{equation}
q^*_{2/3}\, \bar q^*_{2/3} \to W^- q^*_{5/3}\, W^+ \bar q^*_{5/3} \to 
W^- W^+ W^+ b \, W^+ W^- W^- \bar b \, .
\end{equation}
In conclusion, our brief discussion shows that there are characteristic signatures
predicted by these 
 models that will distinguish them from  other extensions of the SM.
While certainly challenging, these signals will be extremely spectacular, and will provide
an  indication of  a new strong dynamics responsible for EWSB.

\subsubsection{Experimental tests of a composite Higgs}

\begin{figure}
\centering\includegraphics[width=.7\linewidth]{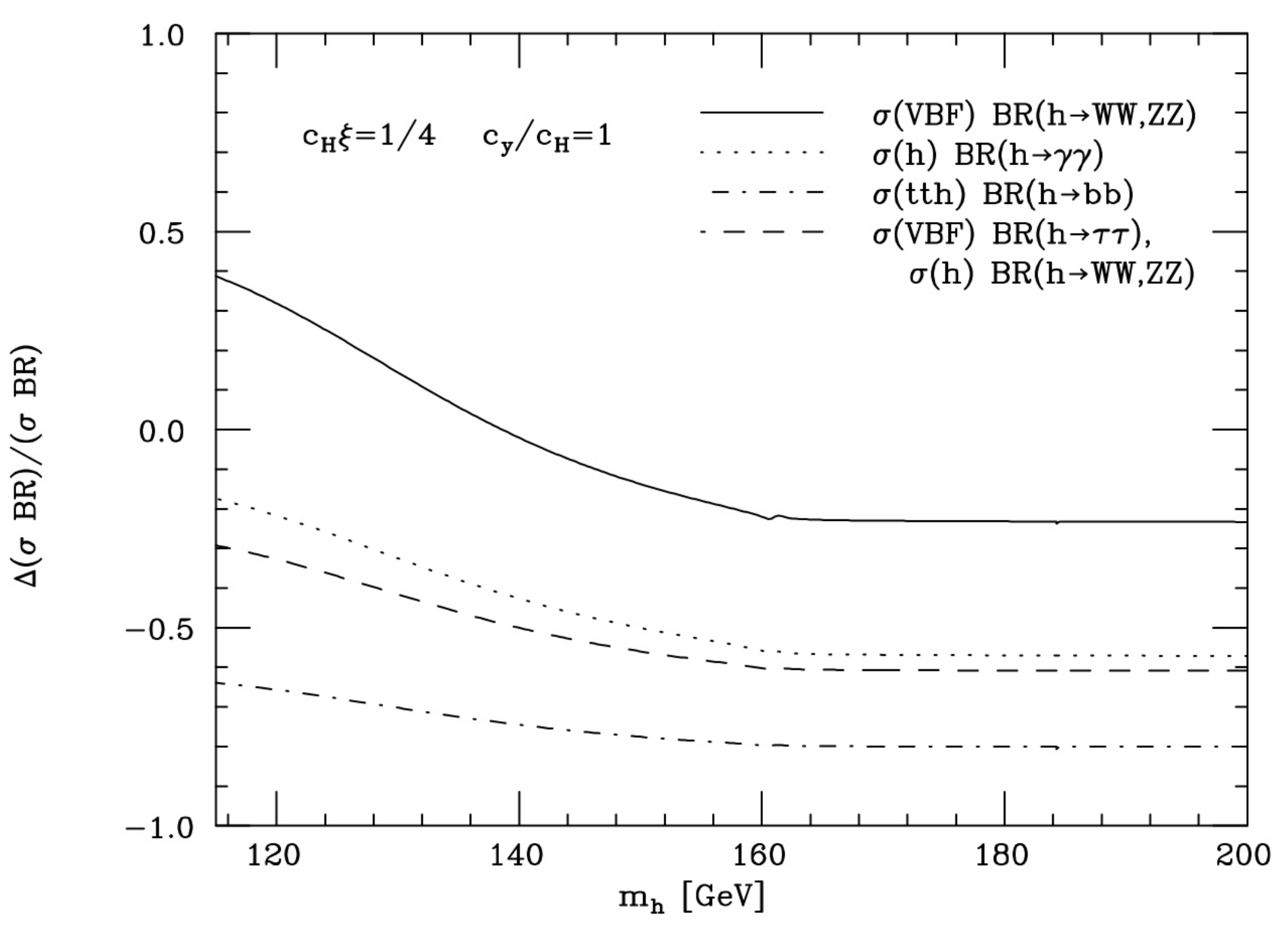}
\caption{The deviations from the SM predictions of Higgs production cross-sections ($\sigma$) and decay branching ratios ($BR$) defined as $\Delta (\sigma ~BR)/(\sigma ~BR)=(\sigma ~BR)_{\rm SILH}/(\sigma ~BR)_{\rm SM} -1$. The predictions are shown for some of the main Higgs discovery channels at the LHC with production via vector-boson fusion (VBF), gluon fusion ($h$), and top-strahlung ($tth$).}
\label{fig1silh}
\end{figure}

As an alternative to the detection of heavy resonances, 
the composite Higgs scenario can also be tested by 
measuring the couplings of the Higgs and   seeing differences  from those of
a SM point-like Higgs.  
For small values of $\xi \equiv v^2/f^2$, as needed to satisfy  the constraint on $\widehat S$,
we can expand  the low-energy Lagrangian in powers of $h/f$
and obtain in this way the following dimension-6 effective Lagrangian 
involving the Higgs doublet $H:$
\begin{eqnarray}
{\cal L}_{\rm SILH} &=&
 \frac{c_H}{2f^2}\partial^\mu \left( H^\dagger H \right) \partial_\mu \left( H^\dagger H \right) 
+ \frac{c_T}{2f^2}\left(H^\dagger {\overleftrightarrow { D^\mu}} H \right)  
\left(   H^\dagger{\overleftrightarrow D}_\mu H\right) \nonumber \\
&-& \frac{c_6\lambda}{f^2}\left( H^\dagger H \right)^3 
+ \left( \frac{c_yy_f}{f^2}H^\dagger H  \bar \psi_L H \psi_R +{\rm h.c.}\right)\, .
\label{lsilh}
\end{eqnarray}
Equation~(\ref{lsilh}) will be referred to as the Strongly Interacting Light Higgs (SILH) 
Lagrangian \cite{silh}.
We have neglected operators suppressed by $1/m_\rho^2$ that are subleading 
versus those of Eq.~(\ref{lsilh}) by a factor $f^2/m^2_\rho\sim N/(16\pi^2)$,
or operators that do not respect the global symmetry $G$ and therefore 
are only induced at the one-loop level with extra suppression factors 
--- see Ref.~\cite{silh}. 
The coefficients $c_H,c_T, c_6$, and $c_y$ are constants of order one that depend on the particular models.
In    5D composite Higgs models they take, at tree-level, the value \cite{silh}:  
$c_H=1$, $c_T=0$, $c_y=1\ (0)$, and $c_6=0\ (1)$  for the  model of Ref.~\cite{cdp} (\cite{acp}).
Only the coefficient  $c_T$ 
 is highly constrained by the experimental data,
 since it contributes to the $\widehat T$-parameter.
 Nevertheless all models with an approximate custodial symmetry 
 give a small contribution to $c_T$.
The other operators  can only be  tested  in Higgs physics. 
They modify the  Higgs decay widths according to
\begin{eqnarray}
\Gamma \left( h\to f\bar f \right)_{\rm SILH} &=& \Gamma \left( h\to f\bar f \right)_{\rm SM} 
\left[ 1-  \xi \left( 2c_y + c_H \right) \right]\nonumber\\
\Gamma \left( h\to WW \right)_{\rm SILH} 
&= &\Gamma \left( h\to WW^{(*)} \right)_{\rm SM} 
\left[ 1- \xi  c_H \right]\nonumber\\
\Gamma \left( h\to ZZ \right)_{\rm SILH} &=& \Gamma \left( h\to ZZ^{(*)} \right)_{\rm SM} 
\left[ 1- \xi c_H  \right]\nonumber\\
\Gamma \left( h\to gg \right)_{\rm SILH} &=& \Gamma \left( h\to gg \right)_{\rm SM} 
\left[ 1- \xi\, {\rm Re}\left( 2c_y + c_H \right) \right]\\
\Gamma \left( h\to \gamma \gamma \right)_{\rm SILH} &=& \Gamma \left( h\to \gamma \gamma \right)_{\rm SM} 
\left[ 1- \xi\, {\rm Re}\left( \frac{2c_y+c_H}{1+J_{\gamma}/I_{\gamma}}+
\frac{c_H}{1+I_{\gamma}/J_{\gamma}}
 \right) \right]\nonumber\\
\Gamma \left( h\to \gamma Z \right)_{\rm SILH} &=& \Gamma \left( h\to \gamma Z \right)_{\rm SM} 
\left[ 1- \xi\, {\rm Re} \left( \frac{2c_y+c_H}{1+J_{Z}/I_{Z}}+
\frac{c_H}{1+I_{Z}/J_{Z}}
 \right) \right]\nonumber\,.
\label{gammas}
\end{eqnarray}
The loop functions $I$ and $J$ are given in Ref.~\cite{silh}.
Note that the contribution from $c_H$ is universal for all Higgs couplings and therefore it does not affect the Higgs branching ratios, but only the total decay width and the production cross-section. The measure of the Higgs decay width at the LHC is very difficult and it can 
only be reasonably done  for a rather heavy Higgs, well above the two gauge boson threshold, that  is not the case of a composite Higgs.
 However, for a light Higgs, LHC experiments can measure the product $\sigma_h \times BR_h$ in many different channels: production through gluon, gauge-boson fusion, and top-strahlung; decay into $b$, $\tau$, $\gamma$ and (virtual) weak gauge bosons.
In Fig.~\ref{fig1silh}, we show the  prediction of a 5D composite Higgs  
for the relative deviation from the SM expectation 
in the main channels for Higgs discovery at the LHC.
At the LHC with about 300~fb$^{-1}$, it will be  possible to measure Higgs production rate times branching ratio in the various channels
 with 20--40~\% precision~\cite{coup}.
 This will translate into a sensitivity on $|c_H \xi |$ and $|c_y \xi |$ up to 0.2--0.4,
 at the edge of the theoretical predictions. 
Since the Higgs coupling determinations at the LHC will  be limited by statistics,  they can benefit from a luminosity upgrade, like the SLHC. At a linear collider, like the ILC, precisions on $\sigma_h \times BR_h$ can reach the per cent level~\cite{ilc}, providing a very sensitive probe on the  scale $f$. 
 
Deviations from the SM predictions of Higgs production and decay rate, could be a hint towards models with strong dynamics.
Nevertheless, they do not unambiguously  imply the existence of a new strong interaction. The most characteristic signals of the SILH Lagrangian have to be found in the very high-energy regime. Indeed, a peculiarity of the SILH Lagrangian
is that, in spite of a light Higgs, longitudinal gauge-boson scattering amplitudes grow with energy and the corresponding interaction can become sizeable. 
 Indeed, the extra Higgs kinetic term proportional to $c_H \xi$ in Eq.~(\ref{lsilh}) prevents Higgs exchange diagrams  from accomplishing the exact cancellation, present in the SM, of the terms growing with energy in the amplitudes. Therefore, although the Higgs is light, we obtain strong $WW$ scattering at high energies.
Using the equivalence theorem~\cite{chan}, it is easy to derive the following high-energy limit of the scattering amplitudes for longitudinal gauge bosons:
\begin{equation}
{\cal A}\left( Z^0_L Z^0_L \to W^+_L W^-_L \right) =
{\cal A}\left( W^+_L W^-_L \to Z^0_L Z^0_L \right)=
-{\cal A}\left( W_L^\pm W_L^\pm \to W_L^\pm W_L^\pm \right)
\nonumber
\end{equation}
\begin{equation}
{\cal A}\left( W_L^\pm W_L^\pm \to W_L^\pm W_L^\pm \right)=-\frac{c_H s}{f^2},
\label{strsc}
\end{equation}
\begin{equation}
{\cal A}\left( W^\pm Z^0_L \to W^\pm Z^0_L\right)=\frac{c_H t}{f^2}\,  ,
{\cal A}\left( W^+_L W^-_L \to W^+_L W^-_L \right) =\frac{c_H (s+t)}{f^2},
\label{strsc2}
\end{equation}
\begin{equation}
{\cal A}\left( Z^0_L Z^0_L \to Z^0_L Z^0_L \right) =0\,.
\label{strsc3}
\end{equation}

\section{Extra dimensions}

One of the  first to postulate the existence of extra dimensions was Kaluza in 1921
\cite{kaluza}.
He wanted to unify gravity with electromagnetism. For this purpose
he considered a 5D theory with only gravity. 
The  5D gravitons, $h_{MN}$ $(M,N=\mu,5)$, corresponds to
fluctuations around the flat space\footnote{Our convention here is $\eta_{MN}=
{\rm Diag}(-1,1,1,1,1)$.} 
$g_{MN}=\eta_{MN}+h_{MN}$. 
From a 
4D point of view, they decompose as a spin-2 particle, $h_{\mu\nu}$, a
spin-1, $h_{\mu 5}$, and spin-zero $h_{55}$. 
Kaluza associated the first one 
with the 4D graviton, and the  second one 
with the photon. This was possible due to the fact that
$
h_{\mu 5}
$
transforms, under an infinitesimal
translation in the extra dimension $y\rightarrow y+\theta(x)$,
as $h_{\mu 5}\rightarrow h_{\mu 5}-\partial_\mu\theta$ 
that corresponds to
a 4D gauge transformation.
Although this 5D theory gave a unified picture of gravity and electromagnetism,
it
failed
 to be realistic since it could not  incorporate massless charged matter.

A second motivation to consider higher-dimensional theories came
from string theory \cite{strings}.
 Strings were found to give  a  consistent
description of  quantum gravity.
Nevertheless, this could only happen if  strings were living
 in more than 4 dimensions.
For example, superstring theory must be formulated in  10 dimensions.
Therefore extra dimensions can be  needed in order to have a consistent
description of quantum gravity.
As we will explain below, it was necessary for these extra dimensions
to be   compactified  with a  compactification radii
 around the Planck length $R\sim 10^{-32}$ cm,
and therefore out of the reach of any experiment.

In 1998, however, Arkani-Hamed, Dimopoulos, and Dvali
 realized that
extra dimensions
could be larger than the Planck length if only gravity was propagating 
in these extra dimensions \cite{add}.
Furthermore, the existence of extra dimensions for gravity
 could also explain
why gravity was much weaker than the other interactions.
The basic idea is very simple.
If gravity propagates in $4+d$ dimensions we know, by Gauss's law,
that the force  between two bodies of masses
$m_1$ and $m_2$ separated by a  distance $r$
is given by
\begin{equation}
F={G_{grav}}\frac{m_1m_2}{r^{2+d}}\, ,
\label{nl}
\end{equation}
where $G_{grav}$ is the equivalent to Newton's constant 
in $4+d$ dimensions.
From Eq.~(\ref{nl}) one learns that 
the gravity force can be weaker than the other gauge forces, not because
the strength of the interaction, $G_{grav}$, is small, but because gravity 
propagates in more than 4D and then 
the gravity force decreases faster  as $r$ increases, 
$F\sim 1/r^{2+d}$, than
the gauge forces, $F\sim 1/r^2$.
Of course, we know that at very large distances 
gravity lives in 4D, since we know that 
Newton's law  reproduces very accurately, for example,
the orbits of the planets.
This means that the extra dimensions must be compact
with a compactification radius $R$.
At distances larger than $R$ we will have a 4D theory with Newton's law:
\begin{equation}
F=G_N\frac{m_1m_2}{r^{2}}\, ,
\label{nl2}
\end{equation}
where $G_N$ is the observed Newton's constant.
Matching Eqs.~(\ref{nl}) and (\ref{nl2})  at $r=R$ one gets
\begin{equation}
  \label{ed}
  G_N=\frac{G_{grav}}{R^d}\, .
\end{equation}
Therefore large $R$ implies a small  $G_N$.
In other words, 4D gravity must be  weaker than the other interactions
if its field lines spread over  large extra
dimensions. 
The larger the extra dimensions, the weaker is gravity.
This is a  very interesting possibility that, as we will see below,
has  spectacular  phenomenological implications.

Several years later  Randall and Sundrum found
a different reason to have extra dimensions \cite{rs}.
If the extra dimensions were curved or `warped', 
gravitons would behave differently 
than gauge bosons and this
could explain their different  couplings to matter.

Below we will discuss these two scenarios in more detail.
Let us first  explain the situation in 
the  old Kaluza--Klein picture.

\section{Kaluza--Klein theories}

As we said before, Kaluza was one of the  first to consider
theories with more than four dimensions in an attempt to unify gravity with
electromagnetism. 
Klein developed this idea in 1926 using a  formalism that 
is usually called Kaluza--Klein reduction \cite{klein}.
Although their initial motivation and ideas do not seem to be viable,
the   formalism that they and others developed is still useful nowadays.
This is the one that  will be considered 
below.

\begin{figure}[htbp]
    \centering
    \setlength\unitlength{.55cm}
    \begin{picture}(10,10)
        \put(0,0){\includegraphics[width=5cm]{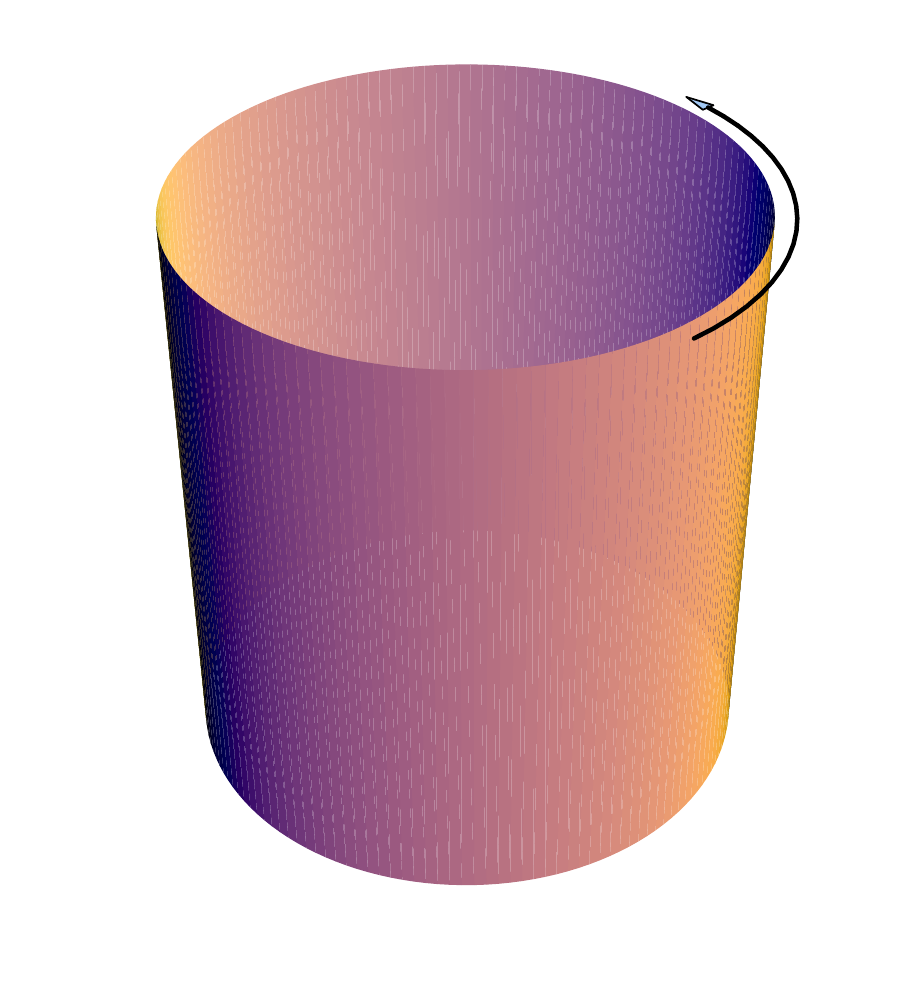}}
        \put(7,6){$y$}
        \put(.7,4){$x_\mu$}
    \end{picture}
    \caption{Compactification on $S^{1}$ }
    \label{circle}
\end{figure}

For simplicity, we will start with a  5D field theory
of scalars.
The action  is given by 
\begin{equation}
  S_5 = -\int d^{4}\! x\, dy\,
M_*\Big[|\partial_\mu\phi|^2+|\partial_y\phi|^2+g^2_5|\phi|^4\Big]\, ,
\label{action}
\end{equation}
where by $y$ we refer to the extra fifth dimension.
We have extracted a universal scale $M_*$ in front of the action in 
order to keep  the 5D field with the same mass-dimension as in 4D.
Let us now consider that the fifth dimension is compact and flat.
We will consider that it has the topology of a circle $S^1$
 as in Fig.~\ref{circle}. This corresponds to the identification of $y$ 
with $y+2\pi R$.
In such a case, we can  expand the 5D complex scalar field in Fourier series:
\begin{equation}
\phi(x,y)=\sum^\infty_{n=-\infty} e^{iny/R}\phi^{(n)}(x)=
\phi^{(0)}(x)+\sum_{n\not= 0} e^{iny/R}\phi^{(n)}(x)\, ,
\label{Fourier}
\end{equation}
that inserted in Eq.~(\ref{action}) and integrated over $y$ gives
\begin{equation}
 S_5 = S_4^{(0)}+S_4^{(n)}\,
\end{equation}
where
\begin{eqnarray}
  S_4^{(0)}&=& -\int d^{4}\! x\,
2\pi R M_*\Big[|\partial_\mu\phi^{(0)}|^2+g^2_5
|\phi^{(0)}|^4\Big]\, ,
\label{action0}\\
S_4^{(n)}&=& -\int d^{4}\! x\,
2\pi R M_*\sum_{n\not= 0}\left[|\partial_\mu\phi^{(n)}|^2+
\left(\frac{n}{R}\right)^2
|\phi^{(n)}|^2\right]+{\rm quartic-couplings}\, .
\label{actionn}
\end{eqnarray}
We  see that the above action corresponds to a 4D theory
with a massless scalar $\phi^{(0)}$  and a tower of massive modes  $\phi^{(n)}$.
The field  $\phi^{(0)}$ will be referred to as the zero-mode, while  $\phi^{(n)}$
will be referred to  as  Kaluza--Klein (KK) modes.

This reduction of a 5D theory to a 4D theory allows one to treat 5D
theories as 4D field theories.
This is very useful since we know much more about 4D theories than 5D theories.
At low energies (large distances) we know that 
massive states in 4D theories
can be neglected.
Therefore the effective theory at energies below $1/R$ is described
by the zero-mode Eq.~(\ref{action0}). After normalizing
$\phi^{(0)}$, we obtain from Eq.~(\ref{action0})
\begin{equation}
 S_4^{(0)} = -\int d^{4}\! x\,
\Big[|\partial_\mu\phi^{(0)}|^2+g^2_4
|\phi^{(0)}|^4 \Big]\, ,
\end{equation}
where the 4D self-coupling is given by
\begin{equation}
g^2_4=\frac{g^2_5}{2\pi RM_*}\, .
\label{strenght}
\end{equation}
This  equation  tells us that
the strength of the interaction of the zero-mode decreases as the
radius increases. 
If $R$ is large, the scalar is weakly coupled.

The general features described above for a 5D scalar 
will also hold for gauge fields and gravity.
After Kaluza--Klein reduction, we will have a 4D theory with a massless
gauge field and a graviton:
\begin{eqnarray}
&& -\int d^{4}\! x\, dy\, M_*\left[ 
\frac{M^2_*}{2} {\cal R}+\frac{1}{4g^2_5} F^{MN}F_{MN}\right]\nonumber\\
&&= -\int d^{4}\! x\, M_*\left[\pi R M_*^2 {\cal R}^{(0)}+
\frac{2\pi R}{4g^2_5}
 F^{(0)\, \mu\nu} F^{(0)}_{\mu\nu}\right]+...\nonumber\\
&&\equiv -\int d^{4}\! x\, \left[\frac{1}{16\pi G_N} {\cal R}^{(0)}+
\frac{1}{4g^2_4}
 F^{(0)\, \mu\nu} F^{(0)}_{\mu\nu}\right]+...\, ,
\label{actionrf}
\end{eqnarray}
where ${\cal R}^{(0)}$ 
is the 4D scalar-curvature containing
the zero-mode (massless) graviton, 
and $ F^{(0)}$ is the gauge field-strength   of the 
zero-mode (massless) gauge boson.
From Eq.~(\ref{actionrf}) we read
\begin{equation}
g^2_4=\frac{g^2_5}{2\pi RM_*}\, ,
\label{gauge}
\end{equation}
for the 4D gauge coupling, and 
\begin{equation}
G_N=\frac{1}{16\pi^2 RM^3_*}\, ,
\label{GN}
\end{equation}
 for the 4D Newton constant.
Again, as in Eq.~(\ref{strenght}), the  strength
of the interaction is suppressed by the length of the extra dimension.

Let us now imagine that we live in 5D. 
From Eqs.~(\ref{gauge})  and (\ref{GN}) 
we learn the following. 
Since the gauge couplings $g^2_4={\cal O}(1)$
and $g^2_5\lesssim 1$ (in order to have a perturbative theory)
we have from  Eq.~(\ref{gauge})  that 
\begin{equation}
R\sim \frac{1}{M_*}\, .
\label{rm}
\end{equation}
On the other hand, 
using the relation $G_N\equiv 1/(8\pi M_P^2)$, where 
$M_P=2.4\times 10^{18}$ GeV is from now on the {\it reduced} Planck scale, we have 
from Eq.~(\ref{GN}) that 
\begin{equation}
M_P^2=2\pi RM^3_*\, .
\label{mp}
\end{equation}
Equations~(\ref{rm}) and (\ref{mp}) imply
\begin{equation}
R\sim \frac{1}{M_P}=l_P\sim 10^{-32}\ {\rm cm}\, .
\label{rmp}
\end{equation}
We have then reached the conclusion that if we live in 5D, 
the radius of the
extra dimension  must be of 
order the Planck length $l_P$!
This extra dimension will not 
be accessible to present or near-future experiments.
This is the reason why experimentalists 
never paid attention to the existence 
of extra dimensions 
even though they were  motivated theoretically a long time ago,
{\it e.g.}, from string theory.

Let us finish this section with a  comment on  the scale $M_*$.
Classically, we introduced this scale based on dimensional grounds.
At the quantum level, however, this scale has a similar meaning as 
$M_P$ in 4D gravity or $1/\sqrt{G_F}$ in Fermi theory. 
It   represents  the cutoff  $\Lambda$  of the 5D theory.
We do not know how to  quantize the 5D theory above $M_*$,
since amplitudes such as 
$\phi\phi\rightarrow \phi\phi$  grow with the energy as  $\sim E/M_*\,$.

\section{Large extra dimensions  for gravity}

In 1998 Arkani-Hamed, Dimopoulos, and Dvali (ADD) proposed 
a different scenario for extra dimensions \cite{add}.
Motivated by the weakness of gravity, they
considered  that only gravity was propagating in the 
extra dimension.
As we already saw,  the effective 5D theory at distances larger than
$R$ is a theory of 4D gravity with a $G_N$ being suppressed by
the length of the extra dimension. 
Then the smallness of $G_N$ can be considered a consequence of 
large extra dimensions. 
The key point to avoid the conclusion
of Eq.~(\ref{rmp}) is that not all fields should share the same dimensions.
In particular, gauge bosons should be localized in a 4D manifold.

In 1995 string theorists realized  that  superstrings in  the strong-coupling limit
contain new solitonic solutions \cite{pol}.
These solutions received 
the name of D-branes and consisted in sub-manifolds of dimensions
D+1 (less than 10) with gauge theories living on them.
From string theory we therefore learn that there can be theories
where gravitons and gauge bosons do not share the same 
number of dimensions, giving realizations of the scenario proposed by 
ADD \cite{aadd}.

Let us then assume that gravity  lives in more dimensions
than the 
SM  particles (leptons, quarks, the Higgs  and gauge bosons),
and study the implications of this scenario.
First of all, we must find out  how large the extra dimensions must be
in order to reproduce the right value of $G_N$.
For $d$ flat and compact extra dimensions, we have
\begin{equation}
 -\int d^{4}\! x\, d^d y\,M^d_*
\frac{M^{2}_*}{2} {\cal R}
= -\int d^{4}\! x\, V^dM^d_* \frac{M^{2}_*}{2}  {\cal R}^{(0)}+...\, ,
\end{equation}
where $V^d$ is the volume of the extra dimensions.
Hence we have 
\begin{equation}
M^2_P=V^dM^{2+d}_*\, .
\label{pref}
\end{equation}
For a toroidal compactification we have $V^d=(2\pi R)^d$.
Following Ref.~\cite{add},
we will absorb the factors $2\pi$ in
$M_*$ and rewrite Eq.~(\ref{pref}) as\footnote{
In string theory, where $M_{st}$ plays the role
of $M_*$,  we have for $d=6$
that $M^2_P=2\pi (RM_{st})^6M^2_{st}/g^4_{4}$.}
\begin{equation}
M^2_P=(RM_*)^dM^2_*\, .
\label{famous}
\end{equation}
Note that Eq.~(\ref{rm}) does not apply 
since gauge bosons do not live in 5D.
Let us  fix $M_*$ slightly above  the electroweak scale $M_*\sim$ TeV to
  avoid introducing a  new  scale  
(this is a nullification of the 
hierarchy problem).
In such a case we have 
from Eq.~(\ref{famous}) a prediction for $R$:
\begin{eqnarray}
d=1\ &\rightarrow&\ R\sim 10^9\ {\rm km}\, ,\nonumber\\
d=2\ &\rightarrow&\ R\sim {\rm 0.5\ mm}\, ,\nonumber\\
\vdots\ \ \ \nonumber\\
d=6\ &\rightarrow&\ R\sim 1/(8\ {\rm MeV})\, ,\nonumber
\end{eqnarray}
The option $d=1$ is clearly ruled out. 
For $d=2$  we expect changes in Newton's law  at distances below the mm.
Surprisingly,  as we will show below, 
we have not measured gravity at distances below $\sim 0.1$~mm.
This is due to the fact that  Van der Waals forces become comparable to gravity at distances
around 1~mm, making it very difficult
to disentangle gravity effects from the large Van der Waals effects.
So the option $d=2$ is being tested today at the present experiments.
Larger values of $d$ are definitely  allowed.

\subsection{Phenomenological implications}

What are the implications of this scenario? 
Let us concentrate on the case $d=2$.
At distances shorter than 1~mm, 
we must notice that gravity lives in 6D. 
To study the effects of a 6D gravity, we will 
again Fourier decompose the 6D graviton field,
$h_{\mu\nu}(x,y_1,y_2)$. For example, if $y_1$ and $y_2$
are compactified  in a torus, we have the Fourier decomposition
\begin{equation}
  \label{eq:kkde}
h_{\mu\nu}(x,y_1,y_2)
=\sum_{n_1=-\infty}^{\infty}\, 
\sum_{n_2=-\infty}^{\infty} e^{i(n_1y_1+n_2y_2)/R}
\ h^{(\vec n)}_{\mu\nu}(x)\, ,
\end{equation}
where $\vec n=(n_1,n_2)$.
The state $h^{(\vec 0)}_{\mu\nu}$ is our massless graviton, 
while $h^{(\vec n)}_{\mu\nu}$ with $\vec n\not=0$
are the   KK states 
that, from a  4D point of view, are massive particles
of masses $m^2_{\vec n}=(n_1^2+n_2^2)/R^2$.
Therefore we can  describe this 6D theory as a  4D theory
containing a massless graviton and a KK tower 
of graviton states.
There are also the KK states for the components $h_{\mu5}$, $h_{\mu6}$,
$h_{65}$, $h_{55}$, and  $h_{66}$.
Nevertheless, 
since matter is assumed to be confined in a 4D manifold at $y=0$,
 we have that the energy-momentum tensor
has only 4D components, $T_{MN}=\eta^{\mu}_M\eta^{\nu}_N
T_{\mu\nu}\delta(y)$.
Hence these extra states do not couple to the 
energy-momentum tensor of matter.
The situation is a little bit more subtle for the `dilaton'
field $\phi$ that corresponds to a 
 combination of  $h_{MN}$ $M,N=5,6$.
Although it does not couple to $T_{\mu\nu}$,
it mixes with the graviton.
This mixing can be eliminated by a Weyl transformation.
Nevertheless, after the Weyl rotation,
$\phi$ appears to be  coupled to the trace of $T_{\mu\nu}$. 
This coupling is usually smaller than those between
gravitons and matter (in fact, it is zero for conformal theories)
and therefore we will neglect it.

The effective Lagrangian for the KK gravitons, 
after normalizing the kinetic term
of the gravitons
 is given 
by 
\begin{equation}
{\cal L}_{KK}= \sum_{\vec n\not=0}\left[{\cal L}_{kin}
-\frac{1}{2}m^2_{\vec n}\left(
 {h^{(-\vec n)}}^{\mu \nu}
  {h^{(\vec n)}}_{\mu \nu}
-{h^{(-\vec n)}}^\mu_\mu {h^{(\vec n)}}^\nu_\nu\right)
+\frac{1}{M_P}h^{(\vec n)\, \mu\nu}
T_{\mu\nu}\right]\, ,
\end{equation}
where ${\cal L}_{kin}$ is the kinetic term of the gravitons. 
The KK states $h^{(\vec n)}_{\mu\nu}$ 
will modify the gravitational interaction at $E>1/R$.
Since they   couple to matter with a strength $\sim 1/M_P$, 
we have that 
at energies $E>1/R$, the (dimensionless)
gravitational strength squared grows as
\begin{equation}
g^2_{grav}\sim\sum_{n_1=0}^{ER}\sum_{n_2=0}^{ER}\frac{E^2}{M_P^2}\sim 
(ER)^2\frac{E^2}{M_P^2}\sim
\left(\frac{E}{M_*}\right)^4\, ,
\label{str}
\end{equation} 
where in the last equality we have used  Eq.~(\ref{famous}).
Note that $g^2_{grav}$ becomes ${\cal O}(1)$ at energies $M_*$.
Therefore $M_*$ is the scale at which quantum gravity effects
are important.
The generalization to $d$ extra dimensions
is given by
\begin{equation}
g^2_{grav}
\sim
\left(\frac{E}{M_*}\right)^{2+d}
\, .
\end{equation} 
With Eq.~(\ref{str}) we can easily estimate any gravitational
effect in any experimental process that we can imagine.

\begin{figure}[htbp]
    \centering
  \includegraphics[width=11cm]{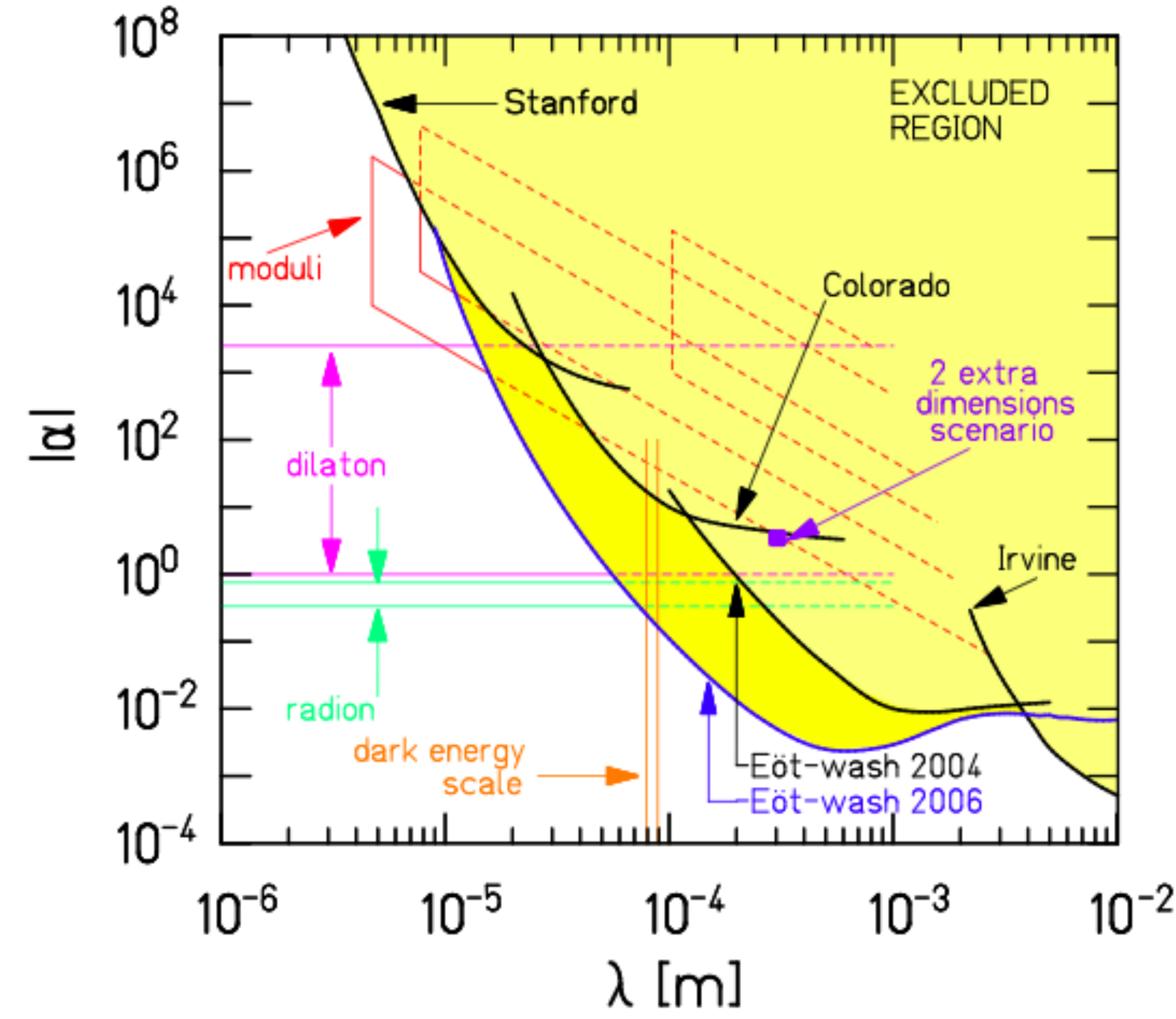}
    \caption{Upper limits on forces of the form of Eq.~(\ref{yukawaforce}) \cite{expgra}} 
    \label{yuex}
\end{figure}

\subsubsection{Measuring the gravitational force 
at millimetre distances}

The KK of the graviton give   rise to new forces.
Since they are massive particles they produce a Yukawa-type force.
For the first   KK ($n_1=\pm 1,n_2=0$ and $n_1=0,n_2=\pm 1$)
of masses $1/R$, this force is given by
\begin{equation}
F_{KK}(r)=-\alpha G_N\frac{m_1m_2}{r}e^{-r/\lambda}\, ,
\label{yukawaforce}
\end{equation}
where $\alpha=16/3$ for a 
2-torus
compactification 
and $\lambda=R$.
Searches for new forces have been carried 
out at several experiments. Nevertheless, the bounds on $\alpha$
are very weak at distances $r$ below  $\sim 0.1$ mm.
In Fig.~\ref{yuex}
 we plot the present experimental bounds on $\alpha$ and $\lambda$.
The value of $R\sim 0.5 $ mm, expected for  $M_*\sim$ TeV and $d=2$,  is  ruled out.
Therefore the case $d=2$ is only at present allowed if
 $M_*\gtrsim 3$ TeV \cite{expgra}.

\subsubsection{Collider experiments}

It is easy to estimate that the  contributions of the KK gravitons
 to any physical process measured in 
any collider   are small.
For example, let us consider the process  $BR(K\rightarrow\pi+gravitons)$.
We can use Eq.~(\ref{str}) with $E\sim M_K$ and obtain
\begin{equation}
BR(K\rightarrow\pi+gravitons)\sim \left(\frac{M_K}{M_*}\right)^4\sim 
10^{-12}\, ,
\end{equation}
for $M_*\sim$ TeV.
This is close to the experimental constraint but it does not rule out
the model.
Similarly, 
we can estimate the  contribution of the KK tower of gravitons to other
low-energy processes:
\begin{eqnarray}
BR(J/\Psi\rightarrow\gamma+gravitons)&\sim& \left(\frac{M_{J/\Psi}}
{M_*}\right)^4\sim 
10^{-10}\, ,\nonumber\\
BR(Z\rightarrow f\bar f+gravitons)&\sim& \left(\frac{M_{Z}}
{M_*}\right)^4\sim 
10^{-4}\, .
\end{eqnarray}
None of them contradict the experimental bounds. 
Until now, no collider experiment has been able to 
exclude this scenario.
The present  limit from colliders arises from the process \cite{grw}
\begin{equation}
qq,gg\rightarrow g+gravitons\, ,
\label{gggrav}
\end{equation}
where the gravitons disappear from the detector carrying energy
with them. 
This cross-section grows with the energy as $E^2/M_*^4$.
Searching for a monojet plus missing transverse energy
one can put a bound on $M_*$.
From Tevatron, one gets $M_*\gtrsim 1$ TeV \cite{tevatronb}.

\subsubsection{Astrophysics}

Will   this scenario modify stellar dynamics?
The KK gravitons can be copiously produced in the stars.
Since they interact very weakly with matter (with 1/$M_P$ suppressed couplings)
they can escape carrying energy with them.
This can definitely change the stellar dynamics.

The most severe constraint on $M_*$ arises from SN 1987A
since it  has a high core temperature $\sim 30$~MeV.
During the collapse of the SN 1987A about 
$10^{53}$ erg
were released in a few seconds.
We must then ensure that the graviton luminosity does not exceed 
 $10^{53}$ erg/s. 

Gravitons can be produced in the supernova core through
several processes.
One example is through nucleon scattering $NN\rightarrow NN+Grav$.
This cross-section can be estimated to be $\sigma\sim \sigma(NN\rightarrow NN)
(E/M_*)^{2+d}$. 
A detailed analysis leads to the bound
$M_*\gtrsim 40,3,1$ TeV for $d=2,3,4$ \cite{astro}. 
A more stringent bound can be found from
KK gravitons emitted by supernova remnants and neutron
stars that are gravitationally trapped,
 forming a
halo, and    occasionally decaying into  photons.
 Limits on $\gamma$-rays
from neutron-star sources imply \cite{Hannestad:2001xi}
 $M_* \gtrsim 200,16$ TeV for
$d = 2,3$. The decay products of the KK gravitons that form the halo
 can provide an extra heat source if they hit the surface of the neutron star;
the low measured luminosities of  pulsars lead to
$M_* \gtrsim 750,35$ TeV for $d= 2,3$. 
Although these bounds tell us that $M_*$ must be larger than the electroweak scale, we must say that they   are  very sensitive to the masses of the lightest KK states that
strongly depend on the type of compactification.

\subsection{Future experiments}

\begin{list}{}{}

\item
{\bf a) Gravity tests at sub-millimetre distances} 

As explained in Ref.~\cite{expgra},
it is difficult to predict how much   future experiments will improve 
on  the  tests of gravity at short-distances,  because the results will almost
surely be limited by systematic errors.
We recall that only for the case $d=2$ does one expect deviations from Newtonian
gravity at accessible distances.

\item
{\bf b) High-energy colliders}

The graviton production Eq.~(\ref{gggrav})  gives a very clean signature at the LHC:
monojet+Missing energy. 
For the LHC with $10$/fb, one expects to probe the model for a 
$M_*$ up to $\sim 8$ TeV. 

Another interesting signature of this scenario is the production of black
 holes \cite{bh}.
In this scenario the Schwarzschild radius is of order
\begin{equation}
R_S\sim \left(\frac{M_{BH}}{M_*}\right)^{\frac{1}{1+d}}\frac{1}{M_*}\, ,
\end{equation}
where $M_{BH}$ is the black hole mass (this is valid only for
$M_{BH}>M_*$).
Estimating the cross-section for the production of black holes as 
$\sigma\sim \pi R^2_S$, we will have  for $M_*\sim$ TeV a production
of $10^{7}$ black holes at the LHC with a luminosity of $30$ fb$^{-1}$.

\end{list}

\section{Warped extra dimensions}

There is another way to escape from the prediction of Eq.~(\ref{rmp})
that does not need to have the SM localized on a 4-dimensional boundary.
This is based on having the extra dimension not flat but `warped'.
This was realized by Randall and Sundrum (RS)~\cite{rs}.
Here we will describe this scenario and will study its phenomenological
consequences. 
Again, as in the ADD scenario,  the motivation is to explain why gravity
is so weak. 

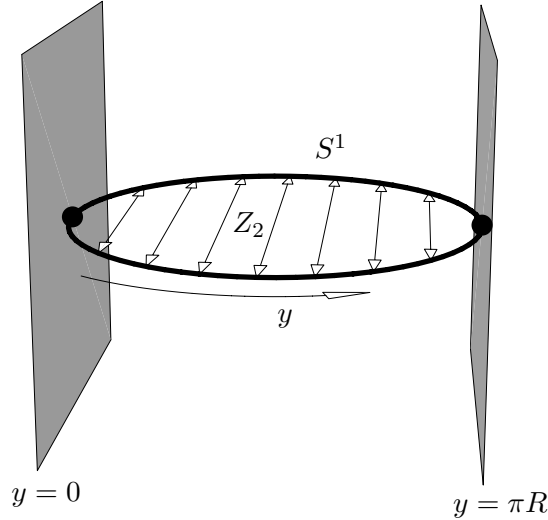
\begin{figure}[htbp]
    \centering
    \setlength\unitlength{.7cm}
    \begin{picture}(10,10)
        \put(0,0){\includegraphics[width=7cm]{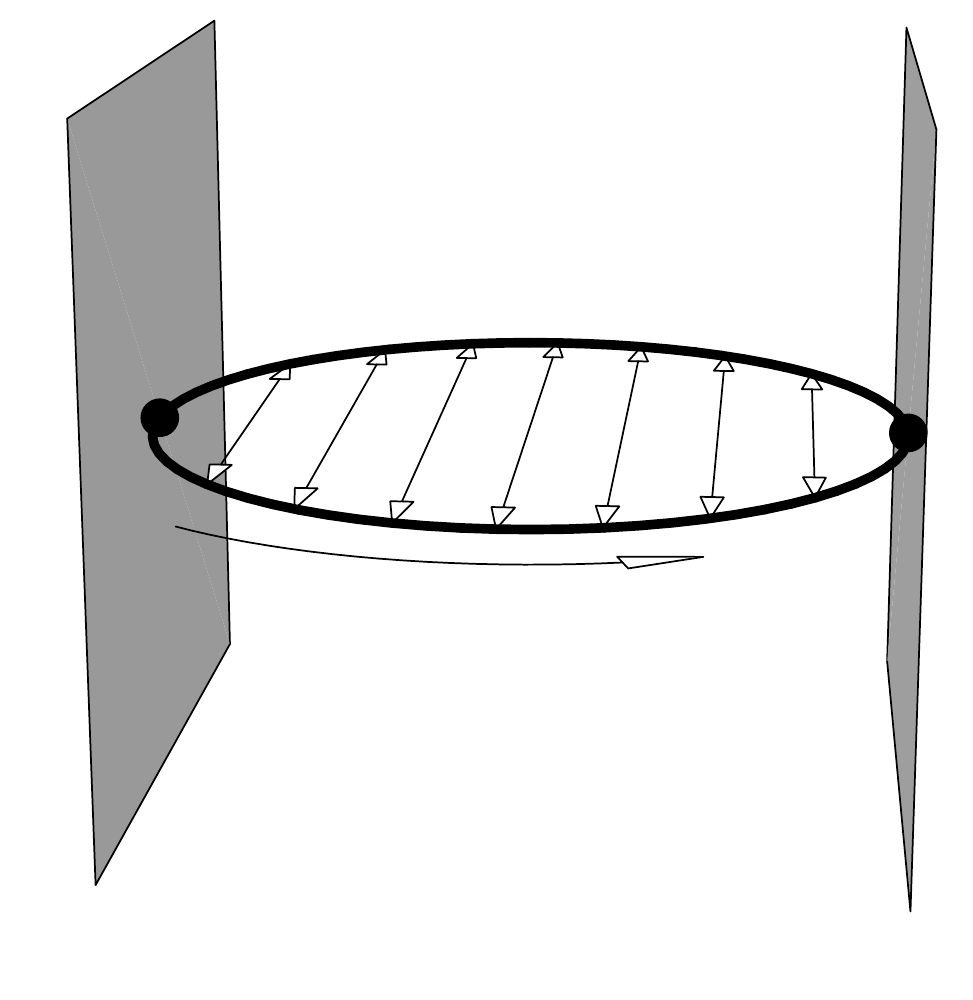}}
        \put(.5,.6){$y=0$} 
        \put(8.8,.4){$y=\pi R$} 
        \put(5.5,4){$y$}
        \put(4.65,5.7){$Z_2$} 
        \put(6.2,7.1){$S^{1}$}
    \end{picture}
    \caption{The  $S^{1}/Z_2$ orbifold}
    \label{fig:orbifold}
\end{figure}

\begin{figure}[htbp]
    \centering
    \setlength\unitlength{.8cm}
    \begin{picture}(10,10)
        \put(0,0){\includegraphics[width=8cm]{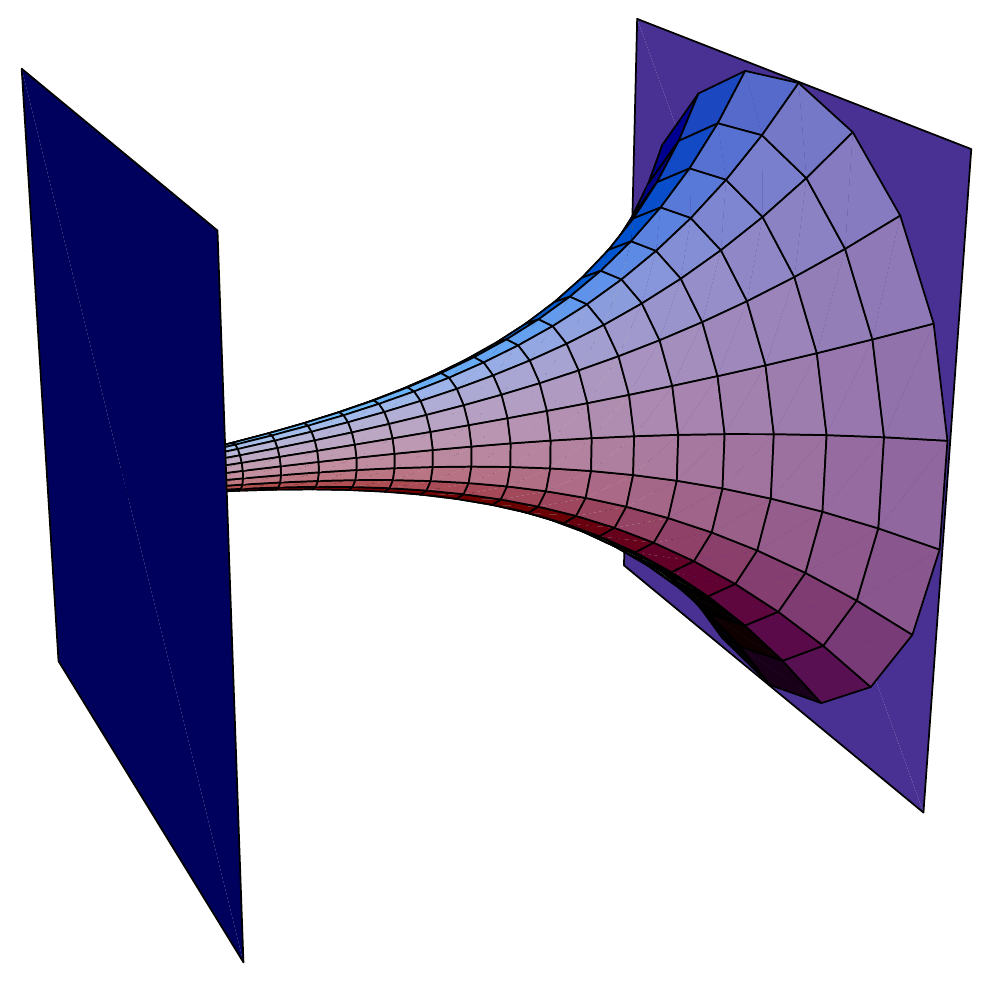}} 
        \put(0,1){$y=0$}
        \put(8.6,1){$y=\pi R$}
        \put(3.4,8.){AdS$_5$}
    \end{picture}
    \caption{The  Randall--Sundrum scenario}
    \label{fig:RS2}
\end{figure}
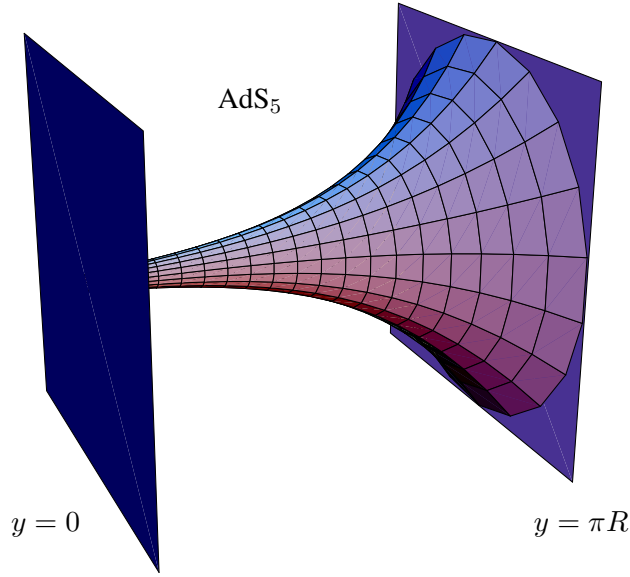

The RS scenario is based on a 5D theory with the extra dimension $y$
compactified in a orbifold, $S^1/Z_2$.   
This compactification 
 corresponds to a circle $S^1$ with the extra identification 
of $y$ with $-y$ as shown in Fig.~\ref{fig:orbifold}.
This gives a `segment' $y\in [0,\pi R]$, a manifold with boundaries\footnote{
This is not a smooth manifold but seems to be a consistent compactification
in string theory.}
at $y=0$ and $y=\pi R$.
Let us now assume that this 5D theory has a cosmological constant in the bulk
 and on the boundaries:
 \begin{equation}
   \label{eq:rs}
S_{RS}=-\int d^{4}\! x\, dy\, \sqrt{-g}
\left[
\frac{1}{2}M^3_*{\cal R}+\Lambda+\delta(y)\Lambda_0+
\delta(y-\pi R)\Lambda_{\pi R}\right]\, .   
 \end{equation}
By solving  Einstein's equations
 \begin{equation}
{\cal R}_{MN}-\frac{1}{2}g_{MN}{\cal R}=-\frac{1}{M^3_*}T_{MN}\, ,
\label{eers}
 \end{equation}
where
 \begin{equation}
-T_{MN}=\Lambda g_{MN}+\Lambda_0\delta(y)g_{\mu\nu}\delta^\mu_M\delta^{\nu}_N
+\Lambda_{\pi R}\delta(y-\pi R)g_{\mu\nu}\delta^\mu_M\delta^{\nu}_N\, ,
 \end{equation}
one obtains the metric
 \begin{equation}
ds^2=e^{2k|y|}dx^\mu dx^\nu \eta_{\mu\nu}+dy^2\, ,
\label{metric}
\end{equation}
where
\begin{equation}
 k=\sqrt{-\frac{\Lambda}{6M^3_*}}\, .
\end{equation}
One must also impose $\Lambda_0=-\Lambda_{\pi R}={\Lambda}/{k}$.
The metric Eq.~(\ref{metric})
 corresponds to a 5D Anti-de-Sitter
(AdS) space (Fig.~\ref{fig:RS2}).  The factor $e^{2k|y|}$
in front of $dx^2$ is called the `warp' factor and 
determines how the 4D scale changes as we move  inside the extra dimension. 
Since $\langle {\cal R}\rangle\propto k^2$,
 we must have $k\lesssim M_*$
in order to be able to use  classical  gravity.

Let us get the effective 4D theory at large distances (below the KK masses).
We have 
 \begin{equation}
S_{RS}=-\int d^{4}\! x\, dy\, \sqrt{-g}
\frac{1}{2}M^3_*{\cal R}=-
\int d^{4}\! x\, \int^{\pi R}_0 dy\, e^{2k y}\frac{1}{2}M^3_*\sqrt{-g^{(0)}}
{\cal R}^{(0)}+...\, ,
\label{actionRS}
\end{equation}
where we have used Eq.~(\ref{metric}) with $\eta_{\mu\nu}\rightarrow
\eta_{\mu\nu}+h_{\mu\nu}^{(0)}$
 (the field 
$h^{(0)}_{\mu\nu}$ is the massless  gravitational
fluctuation about the classical solution Eq.~(\ref{metric})).
Equation~(\ref{actionRS})  implies that the Planck scale
 is given by
 \begin{equation}
M^2_P=\int^{\pi R}_0 dy\, e^{2k y}M^3_*=
\frac{M^3_*}{2k}\left(e^{2k\pi R}-1\right)\, .
\label{mprs}
\end{equation}
We see that the effect of having a warped space is that 
the relation between $M_P$, $M_*$,  and $R$ has changed from
the flat case~[Eq.~(\ref{famous})].
Equation~(\ref{mprs}) is telling us that $M_P$
is exponentially larger than $M_*$. If $M_*$ and $k$ are  fixed to 
be around the TeV,
we can have $M_P\simeq 10^{18}$ GeV for an  extra dimension of radius
\begin{equation}
R\simeq \frac{12}{k}\, .
\label{rrs}
\end{equation}
Hence 
we generated 
a large $M_P$
from a  small  (length of order $1/M_*$) extra dimension.

What about the gauge field \cite{dhr}?
From the 5D action 
\begin{equation}
S_{5}=-\int d^{4}\! x\, dy\, \sqrt{-g}\frac{M_*}{4g^2_5}
F^{MN}F_{MN}\, ,
\end{equation}
we obtain the  effective 4D theory
of the zero-mode gauge field given by
\begin{equation}
S^{(0)}_{5}=-\int d^{4}\! x\, \int^{\pi R}_0 dy\, \sqrt{-g^{(0)}}
\frac{M_*}{4g^2_5}
F^{(0)\, \mu\nu}
F^{(0)}_{\mu\nu}\, .
\label{gaugers}
\end{equation}
Note that no exponential factors appear in front of the $F^2$ term.
Equation~(\ref{gaugers}) tells us that
the 4D gauge coupling is given by
 \begin{equation}
\frac{1}{g^2_4}=\int^{\pi R}_0 dy\, \frac{M_*}{g^2_5}=\frac{\pi RM_*}{g^2_5}\, .
\label{grs}
\end{equation}
This is a very interesting result. It says that $1/g^2_4$ is not
exponentially enhanced as in the case of the graviton.
Therefore the gauge coupling can still be of order 1 
and the model can be
phenomenological viable without the need of localizing the gauge boson
on the 4D boundary.
We see that 
the  warp
 factor 
of the metric Eq.~(\ref{metric}) affects 
fields of different spin in a different way.

There is an alternative way to understand the features of the RS scenario.
If we look at Eq.~(\ref{metric}) 
we see that  the 4D metric $g_{\mu\nu}=e^{2k|y|}\eta_{\mu\nu}$ 
changes as we move in the
extra dimension.
This means that the 4D scales are different in different points of the extra 
dimension. 
As a consequence, for
an observer at $y=0$, 
an experiment
delivering an energy $E$ at $y=0$ will deliver  an energy $Ee^{ky}$
if the experiment is at $y$.
The two energies are related by a blue-shift factor that is the square-root
of the warp factor.
From the point of view of effective theories this means that the 
cutoff of our theory depends on $y$. At $y=0$ this is $M_*$, but
at $y=\pi R$ this is $M_*e^{k\pi R}$.  If we  associate $M_*$  
with the electroweak scale, $M_*\sim$ TeV, the electroweak symmetry breaking
must take place  at $y=0$ (the Higgs must live at $y=0$).
The 4D graviton, however,  must be living at $y=\pi R$ in order to have 
the Planck scale blue-shifted with respect to $M_*$, $M_P\sim M_*e^{k\pi R}$.
This is exactly the situation of the RS scenario (as we will see below, the
massless graviton is localized at $y=\pi R$) giving an intuitive explanation of
Eq.~(\ref{mprs}).

\subsection{KK reduction and phenomenology}

In order to study the full implications of warped extra-dimensions   one must
study the effects of the KK states of the graviton, gauge fields, and fermions.
Here we will perform a KK reduction of the graviton in the background
(\ref{metric}).

First let us decompose the graviton as 
 \cite{jaume}
 $h_{\mu\nu}(x,y)=\sum_n f_n(y)
h^{(n)}_{\mu\nu}(x)$.
Since the metric Eq.~(\ref{metric}) does not depend on $x$,
$h^{(n)}_{\mu\nu}(x)$ corresponds to plane-waves 
$h^{(n)}_{\mu\nu}(x)\propto e^{ip_nx}$ where
$p^2_n=m^2_n$.
From the linearized Einstein equation in the background (\ref{metric})
we obtain\footnote{We take the  gauge $h^\mu_\mu=\partial_\mu h^\mu_\nu=0$.}
\begin{equation}
\left[\partial_y^2-4k^2+e^{-2k|y|}m^2_n-
4k[\delta(y)-\delta(y-\pi R)]\right]f_n(y)=0\, .
\label{kkrs}
\end{equation}
Solving this equation will give us the wave-functions $f_n$
and the masses $m_n$.
We obtain\footnote{We must also impose $f_n(y)=f_n(-y)$ due to the orbifold 
condition.}
\begin{equation}
f_n(y)=\frac{1}{N_n}
   \left[J_{2}(\frac{m_n}{k}e^{-k|y|})+b(m_n)
   Y_{2}(\frac{m_n}{k}e^{-k|y|})\right]\, ,
\end{equation}
where $J_\alpha$ and $Y_\alpha$ are Bessel functions and
$N_n$ are normalization constants.
The values of $b(m_n)$ and $m_n$ are determined by the boundary conditions
that give two equations
\begin{eqnarray}
  b(m_n)&=&-\frac{J_1(\frac{m_n}{k})}
             {Y_1(\frac{m_n}{k})}~,\\
  b(m_n)&=&b(m_n e^{-\pi kR})\label{masscond}~.
\end{eqnarray}
The values of $m_n$ are therefore quantized. 
The lowest mode ($n=0$) corresponds to a massless state $m_0=0$
(that we associate to the 4D graviton). It has a wave-function given by
\begin{equation}
f_0(y)=\frac{e^{2k|y|}}{N_0}\, .
\end{equation}
We see that, as expected, this graviton is localized towards the $y=\pi R$
boundary. The RS scenario corresponds to a theory with localized gravity.
The other KK have masses
\begin{equation}
\label{kkgrav}
     m_n\simeq \left(n+\frac{1}{4}\right) \pi k \, .
\end{equation}
These are  of order of $k\sim$  TeV.
This is very different from the ADD scenario where the KK gravitons 
are very light.
The wave-functions of the KK gravitons, $f_n$, 
are picked towards the boundary at $y=0$.
Then they correspond to states  localized at $y=0$.

The phenomenology of the warped extra-dimensional scenario is  very different from that of ADD.
Since the KK are heavy, there is no implications for low-energy processes.
Only accelerators at very high energies such as the LHC will be able
to test this scenario. The process will be the same as for the ADD scenario,
$qq,gg\rightarrow g+gravitons$ but now only a single  KK state 
will be produced.

If the SM fields propagate in the extra dimension \cite{gepo}  (only the Higgs
must live on the $y=0$ boundary since, as we said above, the
electroweak-breaking sector must be localized at $y=0$),
 KK modes associated to the SM fields  could be seen
in future colliders. 

\subsection{The AdS/CFT correspondence, Higgsless and composite Higgs models}

The AdS/CFT correspondence  relates  5D theories of gravity 
in AdS to  4D strongly-coupled conformal field theories~\cite{adscft}.
In the case of a slice of AdS (Fig.~\ref{fig:RS2}), a similar correspondence
can also be formulated~\cite{adscft2}.
The boundary at $y=\pi R$  
corresponds to an ultraviolet cutoff in the 4D CFT
and to the gauging of certain global symmetries. For example, in the 
case we are considering where gravity and the SM
gauge bosons live in the bulk,
the corresponding 4D CFT will have the Poincar\'{e} group gauged
(giving rise to gravity) and also the SM
group $SU(3)\times SU(2)_L\times U(1)_Y$
(giving rise to the SM gauge bosons).
Matter localized on the boundary at $y=\pi R$
 corresponds to elementary fields  external to the CFT
 that only interact  via gravity and gauge interactions.
On the other hand, the boundary at $y=0$ corresponds in the dual theory 
to an infrared cutoff of the CFT.
In other words, it corresponds to breaking 
the conformal symmetry at the TeV scale.
The KK states of the 5D theory correspond to the bound 
states of the strongly coupled CFT.
Although the CFT picture is useful for understanding some qualitative 
aspects of the theory, it is practically useless for obtaining
quantitative predictions since the theory is strongly coupled.
In this sense, the 5D gravitational theory in a slice of AdS represents
a very useful tool since it allows one to calculate the particle 
spectrum, which would otherwise be unknown from the CFT side.

Following the AdS/CFT correspondence  we can  design five-dimensional  models with the properties
of the strongly-coupled models discussed in Section~\ref{strongsec}.
For example, Higgsless models  \cite{Csaki:2003zu}
 consist in  gauge theories  in  RS spaces   with the symmetry pattern

\begin{equation}
\begin{array}{cc}
 {\rm Boundary\ at\ }y=0\!:      & \qquad  SU(2)_{V}\times U(1)_{X}\times SU(3)_c \\
{\rm 5D\ Bulk}\!:                      & \qquad  SU(2)_L\times SU(2)_R\times U(1)_{X} \times SU(3)_c \\
{\rm Boundary \ at\ }y=\pi R\!: & \qquad  SU(2)_L\times U(1)_Y\times SU(3)_c \\
\end{array}
\end{equation}

\begin{figure}
\centering\includegraphics[width=.7\linewidth]{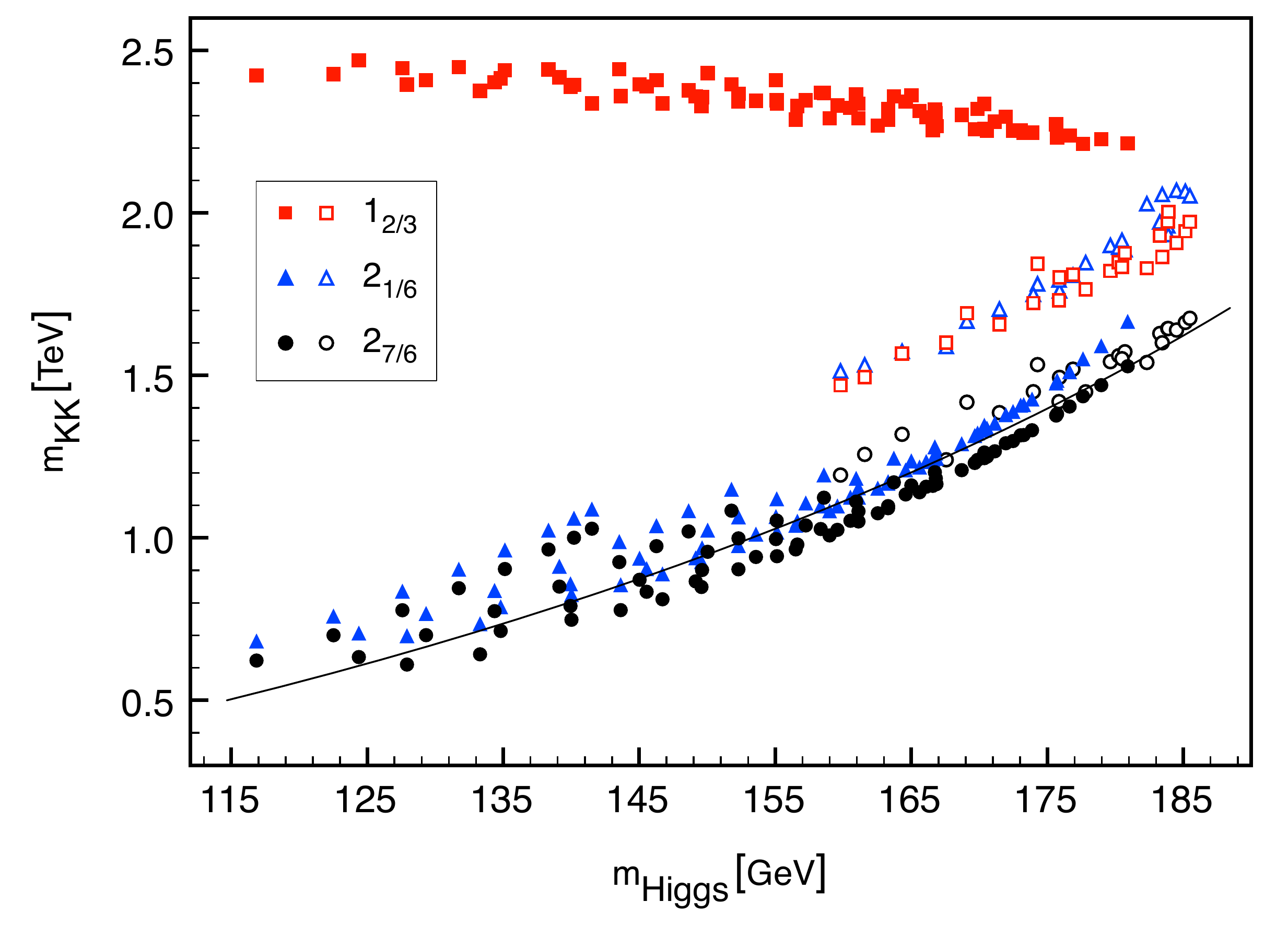}
\caption{KK fermion masses vs the Higgs mass in the model of Ref.~\cite{cdp}.
All fermion KK states are colour triplets under the strong group.
 The quantum numbers  under $SU(2)_L\times U(1)_Y$ are also given. 
We notice that  the  normalization of  hypercharge in Ref.~\cite{cdp} is different from ours; one must  multiply by 2 to get the  hypercharges as defined here. }
\label{higgskk}
\end{figure}

For composite PGB Higgs models we have~\cite{acp,cdp}
\begin{equation}
\begin{array}{cc}
 {\rm Boundary\ at\ }y=0\!: &\qquad O(4)\times U(1)_{X}\times SU(3)_c \\
{\rm 5D\ Bulk}\!:&\qquad  SO(5)\times  U(1)_{X}\times SU(3)_c  \\
 {\rm Boundary \ at\ }y=\pi R\!:&\qquad SU(2)_L\times U(1)_Y\times SU(3)_c \\
\end{array}
\end{equation}
In these models the lightest KK states are the  partners of the top  with SM  quantum numbers ${\bf (3,2)_{7/3,1/3}}$
and $ {\bf (3,1)_{4/3}}$.The spectrum is shown in Fig.~\ref{higgskk}.
Gauge boson and graviton KK states are heavier, around $2.5$ TeV and  $4$ TeV respectively.

\subsection*{Acknowledgements}

I would  like to thank the  organizers  of the 2010 European
School of High-Energy Physics for such a    successful and  stimulating  school.  
This  work  is partly supported by 
CICYT-FEDER-FPA2008-01430, 2009SGR894, UniverseNet (MRTN-CT-2006-035863),
and ICREA Academia program.

\end{document}